\documentstyle[preprint,aps,eqsecnum]{revtex}
\tighten
\begin{document}
% Some general macros
\def\eps{\varepsilon}
\def\klein#1{{\mbox{\scriptsize #1}}}
\def\pzero{{\phantom{0}}}
%\font\bb=cmbxti10
%\def\vec#1{{\bb #1}}

\draft

\title{Brownian motion ensembles\protect\\
 and parametric correlations of the transmission eigenvalues:\protect\\
Application to coupled quantum billiards\protect\\
and to disordered wires.}

\author{Klaus Frahm and Jean-Louis Pichard}
\address{Service de Physique de l'\'Etat condens\'e.
	CEA Saclay, 91191 Gif-sur-Yvette, France.}

\date{\today}

\maketitle

\begin{abstract}
  The parametric correlations of the transmission eigenvalues
$T_i$ of a $N$-channel quantum scatterer are calculated assuming two different
Brownian motion ensembles. The first one is the original ensemble
introduced by Dyson and assumes an isotropic diffusion for the
$S$-matrix. We derive the corresponding Fokker-Planck equation for
the transmission eigenvalues, which can be mapped for the unitary
case onto an exactly solvable problem of $N$ non-interacting fermions
in one dimension with imaginary time. We recover for the $T_i$ the
same universal parametric correlation than the ones recently obtained
for the energy levels, within certain limits. As an application,
we consider transmission through two chaotic cavities weakly
coupled by a $n$-channel point contact when a magnetic field is applied.
The $S$-matrix of each chaotic cavity is assumed to belong to the
Dyson circular unitary ensemble (CUE) and one has a $2\times$ CUE $\to$
one CUE crossover when $n$ increases. We calculate all types
of correlation functions for the transmission eigenvalues $T_i$ and
we get exact finite $N$ results for the averaged conductance
$\langle g\rangle$ and its variance $\langle \delta g^2\rangle$,
as a function of the parameter $n$. The second Brownian motion ensemble
assumes for the transfer matrix $M$ an isotropic diffusion
yielded by a multiplicative combination law. This model is known to
describe a disordered wire of length $L$ and gives another Fokker-Planck
equation which describes the $L$-dependence of the $T_i$.
An exact solution of this equation in the
unitary case has recently been obtained by Beenakker and Rejaei, which
gives their $L$-dependent joint probability distribution.
Using this result, we show how to calculate all types of correlation
functions,
for arbitrary $L$ and $N$. This allows us to get an integral expression
for the average conductance which coincides in the limit $N\to\infty$
with the microscopic non linear $\sigma$-model results obtained by
 Zirnbauer et al, establishing the equivalence of the two approaches.
We review the qualitative differences between transmission through
two weakly coupled
quantum dots and through a disordered line and we discuss the
mathematical analogies between the Fokker-Planck equations of the two
Brownian motion models.
\end{abstract}

\tableofcontents

\section{Introduction}

\label{section:1}

 It is now rather well established that the quantum energy
levels of classically chaotic billiards have Wigner--Dyson
correlations \cite{bohigas2,berry1}. These universal
correlations are closely associated with the concept
of ``quantum chaos'' and have first been obtained in
random matrix theory \cite{mehta} (RMT) in a different context
(statistical description of the spectral
properties of complex nuclei and small metallic particles).
Futhermore, many works\cite{dyson2,dyson3,alt1,alt2,been1,shast,gasp1}
 have considered the case where the
system Hamiltonian $H$ depends on an external parameter $t$.
Instead of a single matrix ensemble defined by
a certain probability density, a continuous family of ensembles,
defined by a $t$--dependent probability density, is considered.
The universality of the level correlations
can then be extended to a broader domain, which involves not only (small)
level separations at a given $t$, but also (small) external parameter
separation $\delta t$. This broader universality has been numerically
checked in very different physical systems: \mbox{e.~g.} disordered
conductors\cite{alt1}, quantum billiards\cite{stone}, the hydrogen atom
in a magnetic field\cite{Del} and
correlated electron systems\cite{alt3,mont1}. Analytical derivations
use perturbation
theory\cite{alt1} or the non linear $\sigma$-model \cite{alt2}.
 Another fruitful way to calculate the
universal expressions of those parametric level correlations is based on
Brownian motion ensembles of random matrices, first introduced by Dyson.
The idea is to assume that the $N \times N$ Hamiltonian matrix
diffuses when one varies the parameter $t$ in the available manifold
given its symmetries.  The $t$-dependence of the level
distribution is then determined by a Fokker--Planck
equation\cite{been1,shast}, which can
be solved in the large $N$-limit for arbitrary symmetries and for
finite $N$ in the unitary case (Sutherland method).

 For quantum transport in the mesoscopic regime, the system is
not closed, but open to electron reservoirs, and becomes
a $N$-channel scatterer with transmission eigenvalues $T_i$.
Many transport properties\cite{jalabert1} (conductance, quantum shot noise...)
are linear statistics of the $N$ eigenvalues $T_i$, which are the
appropriate levels for a scattering problem. The scattering matrix
$S$ or transfer matrix $M$, suitably parametrized by the $N$
transmission eigenvalues $T_i$ and certain auxiliary unitary matrices,
are the relevant matrices for which RMT approaches have been formulated.
For instance, if the unitary $S$-matrix is distributed according
to one of Dyson's classical circular ensembles \cite{dyson1}, the
corresponding transmission eigenvalue distributions have been
recently obtained \cite{jalabert1,baranger1},
exhibiting the same logaritmic pairwise repulsion than for the energy
levels, and hence giving essentially the same Wigner--Dyson correlations
for the transmission eigenvalues.
This RMT description is mainly relevant for quantum
transport through ballistic chaotic cavities, which is a recent
field of significant experimental investigations. Such cavities can
be made with semiconductor nanostructures known as quantum
dots \cite{experiments} with a few channel contacts to two electron
reservoirs. The validity of this RMT description is confirmed by
microscopic semiclassical approaches\cite{bluemel} and by numerical
quantum calculations\cite{baranger1,jalabert3}.

 The universality of the transmission eigenvalue correlations are
established for a given $t$, and therefore we want to know if it can be
extended to their parametric dependence, too, as for the energy levels.
Futhermore, we want to see if these parametric correlations obey a similar
behavior than that of the energy levels. This is one of the
issues which we address in this work, using two Brownian motion
ensembles for the transmission eigenvalues, which result from two
different assumptions concerning the $t$--dependence of $S$
($S$-Brownian motion ensemble) or of $M$ ($M$-Brownian motion ensemble).
This issue has been recently considered also in Refs. \cite{macedo1,andreev1}.

 The $S$-Brownian motion ensemble is simply the original ensemble
introduced by Dyson\cite{dyson2,dyson3} for the scattering matrix $S$
\begin{equation}
\label{smatrix}
S=\left(\begin{array}{cc}
r & t' \\
t & r' \\
\end{array}
\right)\quad.
\end{equation}
 The idea is to assume that $S$ {\em diffuses} in the $S$-matrix space
with respect to a fictitious time $t$ of the Brownian motion,
and converges in the limit $t\to\infty$ to a stationary probability
distribution given by one of the well known circular ensembles:
circular unitary ensemble (CUE, $\beta=2$) when there is no time reversal
symmetry (applied magnetic field); circular orthogonal ensemble
(COE, $\beta=1$) when there are time reversal and spin rotation
symmetries ( no spin--orbit coupling or applied magnetic field) and
circular symplectic ensemble (CSE, $\beta=4$) otherwise. Choosing
an initial probability distribution at $t=0$ of higher symmetry than
the stationary limit, e.g. two independent circular ensembles diffusing
towards a single one, one can perform with the parameter $t$ the crossover
from the initial ensemble to the stationary ensemble.
Actually, the Brownian motion ensembles COE $\to$ CUE,
CSE $\to$ CUE, and $2\times$CUE $\to$ CUE have been investigated in
Ref. \cite{shukla1,shukla2} in terms of the scattering phase shifts
$\theta_i$. Semiclassical and numerical justifications of these models
are given in Ref. \cite{shukla2}. The {\it decisive} difference
between our approach and Refs. \cite{shukla1,shukla2} is that we do not
consider the usual eigenvector--eigenvalue parametrization of $S$, but the
parametrization which explicitly uses the $N$ transmission
eigenvalues $T_i$ of $tt^{\dagger}$ as introduced in Ref. \cite{jalabert1}.
 Since there exists neither a simple mathematical method nor an
intuitive way to relate the $2N$ eigenvalues $e^{i\theta_i}$ of $S$
with the $N$ transmission eigenvalues $T_i$, it is
justified to reconsider the Brownian motion in terms of the latter.
This parametrization is suitable to determine the ``time'' dependence
of the $T_i$, by a Fokker--Planck equation which we map,
using a transformation introduced by Sutherland, onto a Schr\"odinger
equation with imaginary time. In the case of the stationary CUE limit,
one has a problem of
$N$ non interacting fermions which we solve exactly for arbitrary $N$.
One obtains for the corresponding parametric correlations of the
transmission eigenvalues the same universal form than for the
energy levels, after an appropriate rescaling and
{\it within certain limits}.

 A physical application of such a parametric ensemble can for example
be realized by two weakly coupled ballistic (irregular) cavities,
each of them connected to one electron reservoir (Fig. 1). The
crossover between two CUE towards a single CUE is performed by
changing the strength of the coupling from nearly uncoupled
cavities (two CUE) to idealy coupled cavities (one CUE), when a small
constant magnetic field is applied. The transmission eigenvalue
parametrization is very well adapted to this case since the initial
condition is easily realized by the requirement $T_i=0$. Another
application of the Dyson Brownian motion ensemble, concerning the crossover
COE $\to$ CUE in terms of the $T_i$ is presented elsewhere \cite{frahm1}.

 Outside the problem of two weakly coupled quantum dots, we are
also motivated by numerical calculations \cite{jalabert2} showing that
the scattering matrix of a quasi one-dimensional disordered wire
exhibits also some properties of the circular ensembles, in the sense
that the correlation of the scattering phase shifts $\theta_i$ were
found to be approximately described by the universal correlations functions
of these ensembles. For the disordered wire, a crossover from one
circular ensemble towards two independent circular ensembles was observed
as the length of the wire exceeds its localization length.
The appropriate statistical description
\cite{dorokhov,mello2,mello3,review_matrix,lesarcs} of the
transmission eigenvalues of disordered wires is
given by a different Brownian motion ensemble than the original Dyson
model, and results from the multiplicative combination
law of the transfer matrix $M$. This is what we call the $M$-Brownian
motion ensemble which yields another Fokker-Planck equation taking
into account the quasi one-dimensional structure. This Fokker-Planck
equation has been solved by Beenakker and Rejaei\cite{rejaei}
in the unitary case, using
the Sutherland transformation. A remarkable result which they found
is that the transmission eigenvalue pairwise interaction is universal,
in the sense that it does not contain any adjustable parameter. It
coincides with the standard logaritmic RMT repulsion for small
eigenvalue separations, and it is halved for larger separations.
The physical interpretation of this halving is unclear, but must be
somewhat related to the statistical decoupling of the reflection
properties of the two opposite system edges. If it is
right, the Dyson Brownian motion ensemble must also exhibit in the
crossover regime $2\times$CUE $\to 1\times$CUE a somewhat similar pairwise
interaction for weak transmission. It is therefore an interesting idea
to compare those two different Brownian motion ensembles for the
transmission eigenvalues. Let us mention the obvious difference between
them: their initial and stationary limits are inverse of each other.
For the $S$--matrix model, we start from no transmission towards a good
CUE transmission. For the $M$--matrix model on the contrary, the initial
condition corresponds to perfect transmission, and
the system evolves with the time $t$ (here to be identified with
conductor length $L$) to the zero transmission localized limit.

  This paper is composed of four main sections.
First, we introduce the transmission eigenvalue parametrization
of $S$ and define the $S$--Brownian motion ensemble (section
\ref{section:2a}).
The corresponding Fokker-Planck equation is derived in Appendix A.
In section \ref{section:2b}, the Fokker-Planck
equation is solved for an arbitrary initial condition in terms of fermionic
one particle Green's functions, using the method introduced by
Beenakker and Rejaei for the $M$-Brownian motion ensemble.
In section \ref{section:2e}, we show how to recover for the
transmission eigenvalues the same universal parametric correlations than
for the energy levels, after rescaling and within certain limits.

In section \ref{section:2add}, we treat completely the system of two coupled
ballistic cavities by the $S$--Brownian motion ensemble for
the unitary case. The time of the Brownian motion is given
in terms of the numbers of channels in the contacts to the electron
reservoirs ($N$) and in the contact between the two cavities ($n$).
In Section \ref{section:2c}, the crossover $2\times$CUE $\to$ CUE
is considered. The method of orthogonal polynomials,
well known in the framework of random matrix theory \cite{mehta}, is
extended to a more general crossover situation and
yields the correlation functions as determinants (Eq. (\ref{corresult}))
containing a function $K_N(x,y;t)$ given as a sum of Legendre polynomials (Eq.
(\ref{knexp})). This Section contains with Eqs.
(\ref{gaverage}-\ref{gautocorr2}) also exact results for the average and the
fluctuations of the conductance. In Section \ref{section:2d}, we show that
the transmission eigenvalue interaction deviates from the pure RMT
logarithmic interaction, in the limit of weakly coupled cavities.
One can note some interesting similarities with the
pairwise interaction of the $M$-Brownian motion model given
in Ref. \cite{rejaei}.

 The third part of this paper (Section \ref{section:3}) reconsiders
the $M$--Brownian motion ensemble of Ref. \cite{mello2} for
disordered wires in the unitary case. We use the exact expression
of the joint probability distribution found in Ref. \cite{rejaei} and
calculate the corresponding correlation functions, using a method very
similar to the one used in Section \ref{section:2c}. This is possible
because, {\it from a mathematical point of view}, the
Brownian motion models for $S$ and $M$ are very similar.
We get an expression of the $m$-point correlations
as a determinant (Eq. (\ref{mellocorrfunc})) with a function
$K_N(\lambda,\tilde\lambda;t)$ given as a sum and an integral
with Legendre polynomials and Legendre functions
(Eq. (\ref{melloknexplizit})). This allows us to prove the
equivalence between the $M$--Brownian motion ensemble and a more
microscopic approach, based on a non linear $\sigma$-model
formulation and supersymmetry, as far as the behavior of the
first two moments of the
conductance $\langle g \rangle$ and $\langle g^2 \rangle$ is concerned
(section \ref{section:3b}).
We emphasize that the solution of the Brownian motion ensemble is now
complete, contrary to the sigma model approach where one only knows
$\langle g \rangle$ and $\langle g^2 \rangle$ in the large $N$-limit.
In Section \ref{section:3c}, some further simplifications
valid in the localized limit lead to asymptotic expressions
for the density and the two point function for the logarithm of
the transmission eigenvalues.

In Section \ref{section:4}, we review
some essential differences between the transmission properties characterizing
two weakly coupled ballistic chaotic cavities ($S$--Brownian motion
ensemble) on one side and an homogeneous disordered wire
($M$--Brownian motion ensemble) on the other side. We give some concluding
remarks in section \ref{section:5}, notably on the mathematical
similarities between the $S$ and $M$ Brownian motion ensembles.

\section{$S$--Brownian motion ensemble for transmission eigenvalues}

\label{section:2}

\subsection{Parametrization, definition and Fokker--Planck equation}

\label{section:2a}

The scattering matrix $S$ of a system connected to two electron reservoirs
by two $N$-channel contacts can be described in the transmission eigenvalue
parametrization by
\begin{equation}
\label{par}
S=
\left(\begin{array}{cc}
v_1  & 0 \\
0  & u_1 \\
\end{array}\right)
\left(\begin{array}{cc}
\sqrt{1-T} & i\sqrt{T} \\
i\sqrt{T} & \sqrt{1-T}  \\
\end{array}\right)
\left(\begin{array}{cc}
v_2  & 0 \\
0  & u_2 \\
\end{array}\right)
\end{equation}
where $v_{1,2}$, $u_{1,2}$ are $N\times N$ unitary
matrices and $T$ is a diagonal $N\times N$-matrix with real entries
$0\leq T_j\leq 1$. The transmission matrix is $t$, but we mean by
``transmission eigenvalues'' $T_j$ the eigenvalues of $t^\dagger t$.
This parametrization has recently \cite{jalabert1,baranger1}
been used to calculate the conductance properties of a ballistic
cavity, which is known to be suitably described by one of Dyson's
circular ensembles
\cite{dyson1}.
In the unitary case ($\beta=2$) all unitary matrices $v_{1,2}$, $u_{1,2}$ are
independent whereas for the orthogonal ($\beta=1$) and symplectic ($\beta=4$)
symmetry classes, one has $v_2=v_1^D$ and $u_2=u_1^D$ (with
$M^D=M^T$ in the orthogonal case and $M^D=JM^T J^T$, $J=-i\sigma_y$, in
the symplectic case). In the following, the transmission eigenvalues $T_j$
are described by angles $\varphi_j\in[0,\pi/2]$ via $T_j=\sin^2\varphi_j$
and therefore
\begin{equation}
\label{rep}
\left(\begin{array}{cc}
\sqrt{1-T} & i\sqrt{T} \\
i\sqrt{T} & \sqrt{1-T}  \\
\end{array}\right)=
\left(\begin{array}{cc}
\cos\varphi & i\sin\varphi \\
i\sin\varphi & \cos\varphi \\
\end{array}\right)=
\exp\left(\begin{array}{cc}
0 & i\varphi \\
i\varphi & 0 \\
\end{array}\right)
\end{equation}
where $\varphi$ is a diagonal matrix with entries $\varphi_j$. The
invariant measure for $S$ in the parametrization
(\ref{par}) and the change of variables (\ref{rep}) is given by
\cite{jalabert1,baranger1}
\begin{equation}
\label{mes}
\mu(dS)=F_\beta(\varphi)\prod_{j} d\varphi_j\ \mu(du_1)\
\mu(du_2)\ \mu(dv_1)\ \mu(dv_2)\
\end{equation}
with
\begin{equation}
\label{fbeta}
F_\beta(\varphi)=C_\beta\prod_{j<k} \left|\sin^2\varphi_j-
\sin^2\varphi_k\right|^\beta\ \prod_j \left(\sin^{\beta-1} \varphi_j
\ \cos\varphi_j\right)\quad,\quad \beta=1,2,4\quad.
\end{equation}
In the cases $\beta=1,4$ only two products $\mu(du_1)\ \mu(dv_1)$
appear.

The Brownian motion ensemble which we consider (cp. Dyson \cite{dyson2})
describes a unitary random matrix which depends on a fictitious time.
A small change on the time variable $t\to t+\delta t$ yields (in the
unitary case) for $S$ the change:
\begin{equation}
\label{S_dynamics}
S(t+\delta t)=S(t)e^{i\delta X},
\end{equation}
where $\delta X$ is an
infinitesimal hermitian random matrix with averages
\begin{equation}
\label{x_statistic}
\langle\delta X_{ij}\rangle=0\quad,\quad
\langle\delta \bar X_{ij}\delta X_{kl}\rangle=
D\delta t\ \delta_{ik}\delta_{jl}
\end{equation}
and that is independently distributed of $S(t)$.
The diffusion constant $D$ will be precised afterwards, depending on
the considered physical problem. We will only consider the unitary case.
The brownian
motion ensembles for the other cases are more involved and their precise
definition can be found in Ref. \cite{dyson2}. Let $p(\varphi,t)$ be the
joint probability density for the angles $\varphi_j$ of $S$ at the time
$t$, obtained after integration over the unitary matrices $v_{1,2}$,
$u_{1,2}$. This Brownian motion is then characterized by the Fokker-Planck
equation:
\begin{equation}
\label{fp1}
\frac{\partial p(\varphi,t)}{\partial t}=
\frac{D}{4}\sum_j \frac{\partial}{\partial \varphi_j}
\left(F_\beta(\varphi)\frac{\partial}{\partial \varphi_j}
\left(F_\beta(\varphi)^{-1}p(\varphi,t)\right)\right)
\end{equation}
with $F_\beta(\varphi)$ given by Eq. (\ref{fbeta}). In the Appendix
this Fokker-Planck equation is derived for
the unitary case $\beta=2$ which is our main concern.

The details of the derivation in the Appendix show
that the Fokker-Planck equation (\ref{fp1}) remains
valid even if the statistics of the matrix $\delta X$ is arbitrarily
chosen (including the case of a constant matrix $\delta X$)
instead of (\ref{x_statistic}). For this to be true, one then needs
instead of Eq. (\ref{x_statistic}) the non-trivial assumption
that the unitary matrices $v_2$ and $u_2$ that appear in
the parametrization (\ref{par}) of $S(t)$ are for each $t$ uniformly
distributed on the space of $N\times N$ unitary matrices (i.e.
they have always two independent $N\times N$ CUE distributions).
The time step $\delta t$ is
then determined by Eq. (\ref{dm_inform}) of the Appendix.
It is interesting to note that this assumption concerning $u_2$ and $v_2$
is fulfilled at least at $t=0$ for the two applications
concerning the crossovers $2\times$CUE $\to$ CUE or COE $\to$ CUE which
are treated here or in Ref. \cite{frahm1} respectively.
We assume that this holds also for arbitrary $t>0$ and the Brownian motion
model is applicable to rather general physical situations.

\subsection{Solution of the Fokker-Planck equation for arbitrary
initial condition}

\label{section:2b}

We want now to solve the Fokker-Planck equation (\ref{fp1}) for an
arbitrary initial condition $p(\varphi,t=0)=\hat p(\varphi)$
and for $\beta=2$. We set $D=4$ for convenience. We
proceed in a similar way as in Ref. \cite{rejaei} and apply the so-called
Sutherland transformation \cite{sutherland}:
\begin{equation}
\label{trans}
\tilde p(\varphi,t)=\left(F_\beta(\varphi)\right)^{-1/2} p(\varphi,t)
\quad.
\end{equation}
The function $\tilde p(\varphi,t)$ fullfils a Schr\"odinger equation
with imaginary time:
\begin{equation}
\frac{\partial \tilde p(\varphi,t)}{\partial t}=-{\cal H}\tilde
p(\varphi,t)
\end{equation}
where ${\cal H}$ is a many particle Hamilton operator
\begin{equation}
\label{ham0}
{\cal H}=-\sum_j \left(\frac{\partial}{\partial \varphi_j}\right)^2
+V(\varphi)
\end{equation}
with a potential $V(\varphi)$ having the form
\begin{eqnarray}
\nonumber
V(\varphi)&=&-\sum_j \frac{1}{\sin^2(2\varphi_j)}\left(
\sin^2\varphi_j+(\beta-1)(3-\beta)\cos^2\varphi_j\right)\\
\label{pot}
&&+\frac{\beta(\beta-2)}{8}\sum_{j\neq k}
\frac{\sin^2(2\varphi_j)+\sin^2(2\varphi_k)}
{\left(\sin^2\varphi_j-\sin^2\varphi_k\right)^2}+C_N
\end{eqnarray}
and the constant
\begin{equation}
\label{constcn}
C_N=-\frac{1}{4}\beta^2 N-\frac{\beta^2}{3} N(N-1)(N-2)
-\frac{1}{2}\beta(\beta+2)N(N-1)\quad.
\end{equation}
One can see that the unitary case $\beta=2$ is apparently much more
simpler, since the complicated many particle Hamilton operator
reduces in this case to a sum of independent one particle Hamiltonians, i.e.
${\cal H}=\sum_j h(\varphi_j)+C_N$ with
\begin{equation}
\label{ham1}
h(\varphi_j)=-\left(\frac{\partial}{\partial \varphi_j}\right)^2-
\frac{1}{\sin^2(2\varphi_j)}\quad.
\end{equation}
The eigenvalue problem of this operator can be solved using
Legendre polynomials. The properly
normalized eigenfunctions
$\psi_n(\varphi_j)$ with eigenvalues $\eps_n$, $n=0,1,2,\ldots$, are given by
\begin{equation}
\label{eigen1}
\psi_n(\varphi_j)=\sqrt{(1+2n)\sin(2\varphi_j)}\ P_n\bigl(\cos(2\varphi_j)
\bigr)\quad,\quad \eps_n=(1+2n)^2\quad.
\end{equation}
The solution of the Fokker-Planck equation can be expressed by the many
particle fermionic Green's function $G(\tilde\varphi,\varphi;t)$,
as already done in Ref. \cite{rejaei} via
\begin{equation}
\label{sol1}
p(\varphi,t)=\int d^N\tilde\varphi\ \hat p(\tilde\varphi)
\,G(\tilde\varphi,\varphi;t)
\end{equation}
where
\begin{eqnarray}
\label{green0}
G(\tilde\varphi,\varphi;t)&=&\frac{1}{N!} \rho(\tilde\varphi)^{-1}
\ \rho(\varphi)\ \det(g(\varphi_i,\tilde \varphi_j;t))\ e^{-C_N t}\quad,\\
\label{rhodef}
\rho(\varphi)&=&\prod_{i<j} \left(\sin^2\varphi_i-\sin^2\varphi_j\right)
\ \prod_j \sqrt{\sin(2\varphi_j)}\quad,\\
\label{green1}
g(\varphi_i,\tilde \varphi_j;t)&=&\sum_{n=0}^\infty
\psi_n(\varphi_i)\,\psi_n(\tilde\varphi_j)\ e^{-\eps_n t}\quad.
\end{eqnarray}
Eq. (\ref{green1}) describes the one particle Green's function,
defined by
\begin{equation}
\label{green1def}
\left(\frac{\partial}{\partial t}
+h(\varphi_i)\right) g(\varphi_i,\tilde\varphi_j;t)=0
\quad,\quad g(\varphi_i,\tilde\varphi_j;0)=
\delta(\varphi_i-\tilde\varphi_j)\quad.
\end{equation}
For technical reasons, we will now switch to new coordinates
$x_j=\cos(2\varphi_j)\in[-1,1]$ which are just the arguments
of the Legendre polynomials. Let $p(x,t)$ be the probability density
in terms of the $x_j$ taking into account the Jacobian of this
transformation, i.e. $\prod_j (4\sin(2\varphi_j)) p(\varphi,t)\to
p(x,t)$. Eqs. (\ref{sol1}-\ref{green1}) are then replaced by
\begin{eqnarray}
\label{xsol1}
p(x,t)&=&\int d^N\tilde x\ \hat p(\tilde x)
\,G(\tilde x,x;t)\quad,\\
\label{xgreen0}
G(\tilde x,x;t)&=&\frac{1}{N!} \rho(\tilde x)^{-1}
\ \rho(x)\ \det(g(x_i,\tilde x_j;t))\ e^{-C_N t}\quad,\\
\label{xrhodef}
\rho(x)&=&\prod_{i>j} \left(x_i-x_j\right)\quad,\\
\label{xgreen1}
g(x_i,\tilde x_j;t)&=&\sum_{n=0}^\infty
\frac{1}{2}(1+2n)\,P_n(x_i)\,P_n(\tilde x_j)\ e^{-\eps_n t}\quad.
\end{eqnarray}
We have used in (\ref{xsol1}-\ref{xgreen1}) the same symbols $p$,
$\hat p$, $G$, $g$ and $\rho$ for objects which are related with
the corresponding objects of (\ref{sol1}-\ref{green1}) via suitable
transformations including the corresponding jacobians, and
some factors $\sqrt{\sin(2\varphi_j)}$.

One can easily check that $p(x,t)$ in the limit $t\to\infty$
corresponds to the CUE distribution.  For this, we need to consider
only the first $N$ contributions, i.e. $n=0,1,\ldots,N-1$, in the
sum of (\ref{xgreen1}).
The matrix $g(x_i,\tilde x_j;t)$ is then the matrix product of two
matrices $P_n(x_i)$, $P_n(\tilde x_j)$ and a diagonal matrix with
entries $(\frac{1}{2}+n)e^{-\eps_n t}$.
Then the determinant of Eq. (\ref{xgreen0}) becomes
proportional to the two Vandermond determinants $\rho(x)$,
$\rho(\tilde x)$ and
to the factor $\exp(-\sum_{n=0}^{N-1} \eps_n t)$ which is just canceled by
the contribution $e^{-C_N t}$ in Eq. (\ref{xgreen0}). This gives:
\begin{equation}
\label{limit}
%\lim_{t\to\infty} p(x,t)\propto \rho^2(x)\quad,
%\quad const.\sim \int d^N \tilde x\ \hat p(\tilde x)
\lim_{t\to\infty} p(x,t)\propto\rho^2(x)
\cdot\int d^N \tilde x\ \hat p(\tilde x)
\end{equation}
which corresponds to a stationary solution equal to $F_2(\varphi)$,
once the jacobian of the transformation $x_j=\cos(2\varphi_j)$ has been
taken into account. The first correction to the limit (\ref{limit})
is proportional to $\exp(-(\eps_N -\eps_{N-1})t)=\exp(-8Nt)$, giving
the typical time scale $t_c=1/(8N)$ for the Brownian motion to reach
the stationary solution.

\subsection{Universal parametric correlations and their limits}

\label{section:2e}

 The universality of the energy level parametric correlation
has recently attracted a considerable interest. In mesoscopic
quantum physics, this remarkable property obtained after an
appropriate rescaling of the variables, was proven by
diagrammatic methods in Ref. \cite{alt1}. An alternative
derivation of this result
can be directly applied to our problem: Beenakker and Rejaei
have indeed shown \cite{been1} how to recover this universal
behavior from a Brownian motion model for the
Hamiltonian ($H$-Brownian motion ensemble). This derivation
can be easily adapted to the $S$-Brownian motion ensemble and
to the transmission eigenvalue parametric correlation.
As in Refs. \cite{alt1,been1}, we do not
consider here a crossover regime (i.e: $2\times$ CUE $\to$ CUE
or COE $\to$ CUE for instance) but just a CUE $\to$ CUE parametric
dependence, which  reduces to the large $t$-behavior of the crossover
of two decoupled $N \times N$ CUE towards a single $2N \times 2N$
CUE considered in section \ref{section:2add}. Our discussion follows
very closely the derivation of Ref. \cite{been1}, where a
parametric dependence on a quantity $X$ was considered,
related to the Brownian
motion time through the relation $X^2=8Nt=t/t_c$. We measure $t$
in units of $8N=t_c^{-1}$ in order to have a stationary distribution
proportional to $e^{-X^2}$, and we use again the variables
$\varphi$. \footnote{We think that they are the appropriate coordinates
since they have a uniform density in the limit $N\to\infty$.}

Usually, one considers \cite{alt1,been1} two types of correlation
functions which are the density correlation function
\begin{eqnarray}
\nonumber
S(\varphi,X,\varphi',X') & = &
\sum_{i,j}\left\langle(\delta(\varphi-\varphi_i(X))\,
\delta(\varphi'-\varphi_j(X'))\right\rangle\\
\label{ss_corr_def}
&&-\sum_i\left\langle(\delta(\varphi-\varphi_i(X))\right\rangle
\sum_j\left\langle(\delta(\varphi'-\varphi_i(X'))\right\rangle
\end{eqnarray}
and the current correlation function
\begin{equation}
\label{cc_corr_def}
C(\varphi,X,\varphi',X')=
\sum_{i,j}\left\langle(
\dot\varphi_i(X)\,\dot\varphi_j(X')\,
\delta(\varphi-\varphi_i(X))\,
\delta(\varphi'-\varphi_j(X'))\right\rangle\quad.
\end{equation}
For an arbitrary linear statistic (e.~g. conductance, shot noise power,...)
of the form
\begin{equation}
\label{lin_stat}
A(X)=\sum_i a(\varphi_i(X))
\end{equation}
we can calculate the correlator via
\begin{equation}
\label{lin_corr}
\langle \delta A(X)\,\delta A(X')\rangle=
\int_0^{\pi/2}d\varphi\int_0^{\pi/2}d\varphi'
\ a(\varphi)a(\varphi')\,S(\varphi,X,\varphi',X')\quad.
\end{equation}
The two correlation functions (\ref{ss_corr_def}) and
(\ref{cc_corr_def}) are related \cite{alt1,been1} through
\begin{equation}
\label{corr_relation}
\frac{\partial^2}{\partial\varphi\partial\varphi'}\,
C(\varphi,X,\varphi',X')=
\frac{\partial^2}{\partial X\partial X'}\,
S(\varphi,X,\varphi',X')\quad.
\end{equation}

In subsection \ref{section:2c}, we calculate the probability density
$R_{1,1}(x,t;y,t+\tau)$ to find the eigenvalue $x$ at time $t$ and the
eigenvalue $y$ at time $t+\tau$. For a given $\tau$ and in the
limit $t\to\infty$, the particular information of the initial condition
will be lost and we have the density correlation function for the
CUE $\to$ CUE parametric dependence:
\begin{eqnarray}
\label{sn_corr_def}
S_N(\varphi,0,\varphi',X)& = &
\lim_{t\to\infty}\ 4\sin(2\varphi)\,\sin(2\varphi')\\
\nonumber
&&\times\Bigl(R_{1,1}(\cos(2\varphi),t;\,\cos(2\varphi'),t+\tau) -
R_1(\cos(2\varphi);t)\,R_1(\cos(2\varphi');t+\tau)\Bigr)\quad.
\end{eqnarray}
The parameter $X$ is now related with $\tau$ through $X^2=\tau/t_c$
and the sinus-prefactors are just the Jacobians due the
variable transformation $x=\cos(2\varphi)$.
The subscript means that $N$ can have arbitrary values and
that the limit $N\to\infty$ has still not been taken.
Using the results (\ref{corr11res}), (\ref{kn2def}) and
(\ref{hkdef}) derived in the next section, we get directly
the exact expression:
\begin{eqnarray}
\nonumber
S_N(\varphi,0,\varphi',X) & = & 4\sin(2\varphi)\,\sin(2\varphi')\\
\label{sn_corr_result}
&&\times\left(\sum_{n=N}^\infty (n+{\textstyle \frac{1}{2}})\,
P_n(\cos(2\varphi))\,P_n(\cos(2\varphi'))\,e^{-\eps_n\tau}\right)\\
\nonumber
&&\times\left(\sum_{k=0}^{N-1} (k+{\textstyle \frac{1}{2}})\,
P_k(\cos(2\varphi))\,P_k(\cos(2\varphi'))\,e^{+\eps_k\tau}\right)
\end{eqnarray}
where $\eps_n=(1+2n)^2$ is the eigenvalue of the one particle Hamiltonian
$h(\varphi)$ given by (\ref{ham1}). In another work\cite{frahm1}, we
have calculated the function $R_{1,1}$ for the crossover COE $\to$ CUE
which differs
at small $t$ from the result (\ref{corr11res}) we have used here.
However, in the limit $t\to\infty$ both functions become identical and
give the expression (\ref{sn_corr_result}) which has in fact
the same structure than in Eq. (5.27) of Ref. \cite{been1}. One has
just to replace the harmonic oscillator eigenfunctions used in
\cite{been1} by the Legendre polynomials appearing in the $S$-Brownian motion.

First, we consider the limit $N\to\infty$ and differences
$\delta\varphi=\varphi'-\varphi$ of the order $1/N$, in order to
use the large $n$ expansion of the Legendre polynomials:
\begin{equation}
\label{legendre_asym}
P_n(\cos(2\varphi))\simeq\sqrt{\frac{2}{\pi n}}\,\frac{1}{\sqrt{\sin(
2\varphi)}}\ {\textstyle \cos\left((2n+1)\varphi-\frac{\pi}{4}\right)}
\end{equation}
and to replace the sums by integrals. After some smoothing over the
fast oscillating terms, the correlation function
depends only on the difference $\delta \varphi$, i.e.
\begin{eqnarray}
\label{sunend_result}
S(\delta\varphi,X) & = & \lim_{N\to\infty} S_N(\varphi,0,\varphi+
\delta\varphi,X)\\
\nonumber
& = & \rho_0^2\,\int_0^1 ds\int_1^\infty ds'\
\cos(\rho_0 \pi s\,\delta\varphi)\,\cos(\rho_0 \pi s'\,\delta\varphi)
\,\exp\Big(\alpha X^2(s^2-s'^2)\Big)
\end{eqnarray}
where $\rho_0=\frac{2N}{\pi}$ is the average density of the variable
$\varphi$ and $\alpha=N/2=\pi\rho_0/4$. The expression (\ref{sunend_result}),
which is valid in the large $N$- and small $\delta\varphi$-limit,
coincides exactly with the result of Ref. \cite{been1} given by a
Brownian motion for the Hamiltonian $H$ and also with a supersymmetric
calculation
for the unitary case \cite{alt2}. The calculations of the current correlation
function (\ref{cc_corr_def}) and of the correlator (\ref{lin_corr})
are then straightforward and can be found in Ref. \cite{been1}. We note
that (\ref{sunend_result}) yields an algebraic decay for
the correlator of an arbitrary linear statistic as $X\to\infty$
\cite{been1}
\begin{equation}
\label{corr_decay_alg}
\langle \delta A(0)\,\delta A(X)\rangle \sim \frac{1}{X^4}\quad,
\end{equation}
which is also found \cite{macedo1} in a simplified Brownian
motion for the transmission eigenvalues.

However, we cannot directly use the expression (\ref{sunend_result})
to evaluate the integrals in (\ref{lin_corr}), due to the restriction
$\delta\varphi\ll 1$. In fact, we will see in the next section that
the conductance correlator decays exponentially in the $S$-Brownian
motion ensemble. In order to understand this behavior, we consider
again the limit $N\to\infty$, but now without any restriction on
$\varphi$ and $\varphi'$. We replace in the exact
expression (\ref{sn_corr_result}) again the Legendre Polynomials
by the asymptotic formula (\ref{legendre_asym}) but now we keep the
discrete sums instead of replacing them by integrals. One can see that
this step is very crucial: the variable $\varphi$ varies between
$0$ and $\pi/2$ and the correlation function should rather be
expanded in a discrete Fourier-series instead of a continuous
Fourier-integral, being responsible \cite{been1} for the algebraic decay
(\ref{corr_decay_alg}). In the limit $N\to\infty$ at fixed ratio
$\tau/t_c=X^2$, the main contributions comes from the terms with
$n-N\ll N$ and $N-k\ll N$ so that we can linearize the
energy difference $\eps_k-\eps_n=4(1+k+n)(k-n)\simeq 8N(k-n)$. In
addition, all fast oscillating terms of the type
$\cos(4N(\cdots)+\cdots)$ are omited. Then, one of the sums is
easily done and we obtain for
the density correlation function the expression
\begin{equation}
\label{ss_better}
\lim_{N\to\infty} S_N(\varphi,0,\varphi',X)= \tilde S(\varphi+\varphi',X)
+\tilde S(\varphi-\varphi',X)
\end{equation}
where the function $\tilde S(\varphi,X)$ is given as a Fourier-series
\begin{equation}
\label{ss_four}
\tilde S(\varphi,X)=2\sum_{n=1}^\infty S_n(X)\,\cos(2n\varphi)=
\sum_{n=-\infty}^\infty S_n(X)\,e^{2in\varphi}
\end{equation}
with Fourier-coefficients
\begin{equation}
\label{ss_coeff}
S_n(X)=\frac{2|n|}{\beta\pi^2}\,e^{-|n|X^2}\quad,\quad \beta=1,2,4\quad.
\end{equation}
Our derivation accounts of course only for the unitary case $\beta=2$ but
Eq. (\ref{ss_coeff}) is valid for the orthogonal ($\beta=1$) and
symplectic case ($\beta=4$) too \cite{been2}. The derivation
for all $\beta$ can be done by an asymptotic expansion of the
Fokker-Planck equation \cite{been2} similarly as in section 4 of
Ref. \cite{been1} and Eq. (\ref{ss_coeff}) is indeed the
discrete version of Eq. (4.14) in Ref. \cite{been1}.

We are now able to calculate the correlator for the linear
statistics (\ref{lin_stat}). If we expand the function $a(\varphi)$ in
a Fourier-series
\begin{equation}
\label{aa_four}
a(\varphi)=\sum_{n=-\infty}^\infty a_n\ e^{2in\varphi}\quad,\quad
a_n=a_{-n}\quad,
\end{equation}
we find
\begin{equation}
\label{lin_corr_better}
\langle \delta A(0)\,\delta A(X)\rangle =\pi^2\sum_{n=1}^\infty a_n^2\,
S_n(X)=\frac{2}{\beta}\sum_{n=1}^\infty a_n^2\,n\,e^{-nX^2}\quad.
\end{equation}
In (\ref{aa_four}), we have extended $a(\varphi)$ to an even
function on the interval $-\pi/2\le \varphi\le\pi/2$ so that the
Fourier-coefficients are even in $n$ (One should imagine that $a$ is
a well defined function of $x=\cos(2\varphi)$.).

As an example we consider the conductance $g=A$ with $a(\varphi)=
\sin^2(\varphi)$, i.e. $a_0=1/2$, $a_{\pm 1}=-1/4$ and
$a_n=0$ if $|n|\ge 2$,
\begin{equation}
\label{gg_corr}
\langle \delta g(0)\,\delta g(X)\rangle =\frac{1}{8\beta}\,e^{-X^2}
=\frac{1}{8\beta}\,e^{-\tau/t_c}
\end{equation}
which is (in the limit $t\to\infty$) consistent with Eq. (\ref{gautocorr2})
of the next section.

The current correlation function is calculated from (\ref{corr_relation})
and the assumption that $S$ and $C$ depend only on the difference
$X'-X$. We obtain
\begin{equation}
\label{cc_better}
C(\varphi,0,\varphi',X)= -\tilde C(\varphi+\varphi',X)
+\tilde C(\varphi-\varphi',X)
\end{equation}
where $\tilde C(\varphi,X)$ has the Fourier-coefficients
\begin{equation}
\label{cc_four}
C_n(X)=-\frac{1}{4n^2}\frac{\partial^2}{\partial X^2}\,S_n(X)
=\frac{1}{\beta\pi^2}\left(1-2|n|X^2\right)\,e^{-|n|X^2}
\quad,\quad n\neq 0\quad.
\end{equation}
The coefficient $C_0(X)$ is not determined and not needed due to
(\ref{cc_better}). The modified sign in the first contribution
in (\ref{cc_better}) is important because the current correlation
function is odd as a function of $\varphi$ (or $\varphi'$)
if $\varphi'$ (or $\varphi$) is fixed.

 In summary, we have found that the parametric correlations of the
$S$-Brownian motion are only {\it locally}
(on a range $\delta\varphi=\varphi'-\varphi\ll 1$) identical with
the universal correlations found by a $H$-Brownian motion
\cite{been1} or by a microscopic approach in the framework of
the zero-dimensional supersymmetric non linear $\sigma$-model
\cite{alt2}. This agreement does not hold if we consider the whole
range of $\varphi$, $\varphi'$. In particular, quantities {\it integrated
over the whole spectrum} like the correlator of a linear statistic
must be treated with more care and
indeed the conductance correlator shows an exponential instead
of an algebraic decay, {\it in the $S$-Brownian motion ensemble}.

\section{Application to two weakly coupled chaotic dots.}

\label{section:2add}

 The $S$-Brownian motion ensemble with CUE stationary limit can
be applied to a system made by two quantum dots in series under
an applied magnetic field, for describing how its transmission
eigenvalue distribution varies as a function
of the strength of their coupling.
We assume that the two cavities are coupled between them by a wire
with $n$ conducting channels (see Fig. 1), and each of them to one
electron reservoir through $N(\neq n)$ channels. The underlying
classical dynamics is supposed to be fully chaotic in each dots.

 For this particular application, we need to express the fictitious time
$t$ of the Brownian motion, which has been introduced in a rather
abstract mathematical way, in terms of the physical parameters $N$
and $n$. In Ref. \cite{bohigas1},
it was shown that a closed system where the classical motion remains
(nearly) in several separated chaotic regions of the classical phase
space can be modelled by a random matrix approach similar to Dyson's
Gaussian Brownian motion ensemble discussed in Ref. \cite{dyson2}.
The time parameter of the latter can then be related with the particle
flux between the nearly separated regions of phase space. In principle,
one can relate \cite{frahm1} the $H$-Brownian motion (for a closed system)
with the $S$-Brownian motion (for an open system)
by some microscopic assumptions about the $S(H)$-dependence and
then apply the results of Ref. \cite{bohigas1} in order to obtain
the physical meaning of the $S$-Brownian motion time.

For the scope of this paper, we will only give a simple heuristic
derivation. Therefore, we assume that the $(N+n)$-dimensional
$S$-matrix of each cavity has a CUE distribution when an external parameter
is varied (Fermi energy, magnetic field, shape of the dot...) and has
the form shown in
Eq. (\ref{smatrix}), excepted that $r$ is $N\times N$-matrix whereas
the coupling gives rise to a $n\times N$-matrix $t$. When one electron
is injected from a reservoir into one of the cavities by the $i$-th of
the $N$ incoming channels of the corresponding lead, one can assume that
the probability that it goes directly back to reservoir is just
\begin{equation}
\label{leftprob}
p_{\mbox{\scriptsize\,out}}=\sum_{j=1}^N \left\langle |r_{ji}|^2
\right\rangle=\frac{N}{N+n}
\end{equation}
and that the probability to go through the $n$ channel wire into the
other cavity is
\begin{equation}
\label{rightprob}
p_{\mbox{\scriptsize\,trans}}=\sum_{j=1}^n \left\langle |t_{ji}|^2
\right\rangle=\frac{n}{N+n}=1-p_{\mbox{\scriptsize\,out}}\quad.
\end{equation}
This results from the assumption that the chaotic motion inside the dot
uniformly explores the dot boundaries. Furthermore,
one can assume that the typical time for the electron to stay in one
cavity is very long compared to the time needed to cross the connection,
i.e. the electron looses completely the memory from which reservoir it was
injected to the actual cavity. In this case, the total probability
$p_{\mbox{\scriptsize\,total}}$ for the
electron to be transmitted from one reservoir to the other is given by
$p_{\mbox{\scriptsize\,total}}=p_{\mbox{\scriptsize\,trans}}(
p_{\mbox{\scriptsize\,out}}+p_{\mbox{\scriptsize\,trans}}\,
p_{\mbox{\scriptsize\,total}})$, a relation which accounts for all
type of processes where the electron can move between the cavities an
arbitrary number of times before it leaves the system into the other
reservoir. The solution of this equation is given by
\begin{equation}
\label{totalprob}
p_{\mbox{\scriptsize\,total}}=\frac{p_{\mbox{\scriptsize\,trans}}}
{1+p_{\mbox{\scriptsize\,trans}}}=\frac{n}{N+2n}\quad.
\end{equation}
Then, the average conductance can be estimated by $\langle g\rangle =N\,
p_{\mbox{\scriptsize\,total}}$ which leads with the exact result
Eq. (\ref{gaverage}) of Section \ref{section:2c} to the identification
\begin{equation}
\label{timemeaning}
t\simeq\frac{1}{8N}\ln\left(1+\frac{2n}{N}\right)
\end{equation}
where $t$ the Brownian motion time. The prefactor in
(\ref{timemeaning}) accounts for
the typical time scale $t_c= 1/(8N)$ which
is found in the end of Section \ref{section:2b}. In the weak
coupling limit $n\ll N$, we have the simple proportionality
$t/t_c\simeq (2n)/N$.

\subsection{Crossover of two CUEs to one CUE for the transmission eigenvalues}

\label{section:2c}

In the parametrization (\ref{par}), the case of two independent
CUEs of dimension $N$ corresponds to
the requirement that all $T_j=0$ or $x_j=1$ respectively. We have
therefore $\hat p(x)=\prod_j \delta(1-x_j)$ for the initial condition.
The evaluation of (\ref{xsol1}) leads to the limit
$p(x,t)=\lim_{\tilde x_j\to 1} G(\tilde x,x;t)$.
We note that the Legendre polynomials are particular
Hypergeometric functions, i.e. $P_n(x)=
F(-n,n+1;1;\frac{1}{2}(1-x))$ which can be expanded as
\begin{equation}
\label{legexp}
P_n(x)=\sum_k q_k\left((n+{\textstyle \frac{1}{2})^2}\right)\ (x-1)^k
\end{equation}
with the polynomials
\begin{equation}
q_k(u)=\frac{1}{k!^2\,2^k}\prod_{\nu=0}^{k-1}
\left(u-(\nu+{\textstyle \frac{1}{2})^2}\right)
\end{equation}
of degree $k$. We now expand the one particle propagator
$g(x_i,\tilde x_j;t)$ with respect to $(1-\tilde x_j)$ up to terms
$(1-\tilde x_j)^{N-1}$. Again we can identify a matrix product
in the determinant of Eq. (\ref{xsol1}) which yields a factor
$\rho(\tilde x)$ just cancelling the inverse contribution in
(\ref{xsol1}). For the further calculation we use the notation
$P_n(x,t)=P_n(x)\,e^{\eps_n t}$ and find
from (\ref{xsol1}), (\ref{xgreen0})
\begin{equation}
\label{prob0}
p(x,t)\propto \det(P_n(x_i,t))\,\det(g_k(x_i,t))
\end{equation}
with
\begin{equation}
\label{gkdef}
g_k(x,t)=\sum_{n=0}^\infty
(n+{\textstyle \frac{1}{2}})\,q_k\left(
(n+{\textstyle \frac{1}{2}})^2\right)\,P_n(x)\,e^{-\eps_n t}\quad.
\end{equation}
Because of the determinant in Eq. (\ref{prob0}), we note that
one can replace in Eq. (\ref{gkdef}) the polynomials
$q_k(u)$, $k=0,1,\ldots,N-1\ $ by an {\em arbitrary\/}
set of linear independent polynomials of (maximal) degree $N-1$.
Such a substitution affects only the normalization constant. In the following,
we will use two different substitutions for two different
purposes: first to calculate exactly all kind of correlation
functions and second to determine the effective {\it interaction}
between the $x_j$ (or $\varphi_j$) variables. We begin with the calculation
of the correlation functions defined by
\begin{equation}
\label{corfunc0}
R_k(x_1,\ldots,x_k;t)=\frac{N!}{(N-k)!}\int dx_{k+1}\cdots dx_N
\ p(x,t)\quad,\quad k=1,2,\ldots,N\quad.
\end{equation}
We put $u_l=(l+\frac{1}{2})^2=\frac{1}{4}\eps_l$,
$l=0,1,\ldots,N-1\ $ and replace the
polynomials $q_k(u)$ by the Lagrangian interpolation polynomials $L_k(u)$ for
the $u_k$, i.e.
\begin{equation}
\label{lagdef}
L_k(u)=\prod_{l=0,\,l\neq k}^{N-1} \frac{u-u_l}{u_k-u_l}
\quad\Rightarrow\quad L_k(u_l)=\delta_{kl}
\end{equation}
where $l,k=0,1,\ldots,N-1$.
We find now for the joint probability density the result
\begin{equation}
\label{prob1}
p(x,t)=\frac{1}{N!}\det(P_n(x_i,t))\,\det(h_k(x_i,t))
\end{equation}
with
\begin{equation}
\label{hkdef}
h_k(x,t)=(k+{\textstyle \frac{1}{2}})\,P_k(x)\,e^{-\eps_k t}+
\sum_{n=N}^\infty (n+{\textstyle \frac{1}{2}})\,
L_k\left((n+{\textstyle \frac{1}{2}})^2\right)\,
P_n(x)\,e^{-\eps_n t}\quad.
\end{equation}
The normalization in (\ref{prob1}) follows directly from the
obvious property
\begin{equation}
\label{quasiorth}
\int_{-1}^1 dx\,P_n(x,t)\,h_k(x,t)=\delta_{nk}\quad.
\end{equation}
Actually, this ``orthogonality'' relation enables us to calculate
all correlation functions (\ref{corfunc0}). We define the function
\begin{equation}
\label{kndef}
K_N(x,y;t)=\sum_{n=0}^{N-1} P_n(x,t)\,h_n(y,t)
\end{equation}
and find the following three properties
\begin{eqnarray}
\label{prob2}
p(x,t)&=&\frac{1}{N!}\det\left(K_N(x_i,x_j;t)_{\,1\leq i,j\leq N}\right)
\quad,\\
\label{knnorm}
\int_{-1}^{1} dx\ K_N(x,x;t)&=&N\quad,\\
\label{knkonv}
\int_{-1}^1 dy\ K_N(x,y;t)\,K_N(y,z;t)&=&K_N(x,z;t)\quad.
\end{eqnarray}
{}From these properties and a well known theorem in the
theory of random matrices \cite{mehta,dyson4} (cp. Theorem 5.2.1. of
Ref. \cite{mehta}),
we find directly the result for the correlation functions
\begin{equation}
\label{corresult}
R_k(x_1,\ldots,x_k;t)=\det\left(K_N(x_i,x_j;t)_{\,1\leq i,j\leq k}\right)
\quad.
\end{equation}
For the cases $k=1,2$, we have
\begin{eqnarray}
\label{corr1}
R_1(x;t) & = & K_N(x,x;t)\quad,\\
\label{corr2}
R_2(x,y;t) & = & R_1(x;t)\,R_1(y;t)-K_N(x,y;t)\,K_N(y,x;t)\quad.
\end{eqnarray}
For practical applications, we give the more explicit expression for
(\ref{kndef})
\begin{equation}
\label{knexp}
K_N(x,y;t)=\sum_{k=0}^{N-1}\frac{1}{2}(1+2k)\,P_k(x)\,P_k(y)+
\sum_{k=0}^{N-1}\sum_{n=N}^\infty (n+{\textstyle \frac{1}{2}})\,
D_{kn}\,P_k(x)\,P_n(y)\,e^{(\eps_k-\eps_n)t}
\end{equation}
with coefficients
\begin{eqnarray}
\label{dkoeffdef}
D_{kn}&=&L_k\left((n+{\textstyle \frac{1}{2}})^2\right)=
\prod_{l=0,\,l\neq k}^{N-1} \frac{(n+l+1)(n-l)}{(k+l+1)(k-l)}\\
&=&\frac{(-1)^{N-1-k} (n+N)!}{(n-N)!(k+N)! (N-k)!}\cdot
\frac{(2k+1)(N-k)}{(n+k+1)(n-k)}
\quad.
\end{eqnarray}
One might also consider correlation functions at two different times, i.e.
\begin{eqnarray}
\nonumber
&&R_{k,l}(x_1,\ldots,x_k,t;y_1,\ldots,y_l;t+\tau)=
\frac{N!^2}{(N-k)!(N-l)!}\\
\label{cordiffdef}
&&\qquad\times
\int dx_{k+1}\cdots dx_N\,
\int dy_{l+1}\cdots dy_N\ p(x,t)\ G(x,y;\tau)\quad.
\end{eqnarray}
Without going into technical details, we mention that one can calculate
the function $R_{1,1}(x,t;y,t)$ with the technic of functional
derivatives \cite{mehta}. The result is
\begin{equation}
\label{corr11res}
R_{1,1}(x,t;y,t+\tau)=R_1(x;t)\,R_1(y;t+\tau)+
\Bigl(g(x,y;\tau)-K_N(x,t;y;t+\tau)\Bigr)\,K_N(y,t+\tau;x,t)
\end{equation}
where
\begin{equation}
\label{kn2def}
K_N(x,t_1;y;t_2)=\sum_{k=0}^{N-1} P_k(x,t_1)\,h_k(y,t_2)
\end{equation}
is a generalization of (\ref{kndef}) for different times. The limit
$\tau\to 0$ is given by $R_{1,1}(x,t;y,t)=R_2(x,y;t)+\delta(x-y)\,R_1(x)$.

The knowledge of the correlations functions ( \ref{corr1}, \ref{corr11res})
enables us to calculate exact expressions for the average of the
conductance $g=\sum_j T_j=\frac{1}{2}\sum_j (1-x_j)$ and the auto correlation
of its variance $\delta g=g-\langle g\rangle$. Using the orthogonality of
the Legendre polynomials and their recursion relation we find
\begin{eqnarray}
\label{gaverage}
\langle g(t) \rangle &=&\frac{N}{2}\left(1-e^{-8Nt}\right)\quad,\\
\nonumber
\langle \delta g(t)\,\delta g(t+\tau) \rangle
&=&\frac{N^2}{4}\,e^{-8N\tau}\,\biggl\{\frac{1}{4N^2-1}+
\frac{N+1}{2N+1}\,e^{-8(2N+1)t}\\
\label{gautocorr}
&&\qquad\qquad\quad+\frac{N-1}{2N-1}\,e^{-8(2N-1)t}\,-1
\cdot e^{-16Nt}\biggr\}\quad.
\end{eqnarray}
Of particular interest is the large $N$ limit. In order to obtain
a non trivial crossover, we have to measure the time variable
in units of the typical crossover scale $t_c=1/(8N)$ and
to keep the ratio $t/t_c$ fixed in the limit $N\gg 1$.
We find then up to corrections of order $1/N$
\begin{eqnarray}
\label{gaverage2}
\langle g(t) \rangle &=&\frac{N}{2}\left(1-e^{-t/t_c}\right)\quad,\\
\label{gautocorr2}
\langle \delta g(t)\,\delta g(t+\tau) \rangle
&=&\frac{1}{16}\,e^{-\tau/t_c}\,\left\{1+\left(\frac{2t^2}{t_c^2}
-\frac{2t}{t_c}-1\right)
e^{-2t/t_c}\right\}\quad.
\end{eqnarray}
%In the limit of two independent CUE ensembles, i.e. a small connection
%of two ballistic cavities, we find
%\begin{equation}
%\label{gsmallt}
%\langle g(\tilde t) \rangle\simeq \frac{N}{2}\tilde t\quad,
%\quad \langle \delta g^2(\tilde t)\rangle\simeq \frac{1}{4}\tilde t^2
%\simeq \frac{1}{N^2}\langle g(\tilde t) \rangle^2
%\quad.
%\end{equation}

\subsection{Effective pairwise interaction in the crossover regime}

\label{section:2d}

We want evaluate (\ref{prob0}) in a suitable way for
having the effective pairwise interaction between the $x_j$- or
$\varphi_j$-variables. We now replace in Eq. (\ref{gkdef}) the polynomials
$q_k\left((n+\frac{1}{2})^{2}\right)$ by $4^k(n+\frac{1}{2})^{2k}=\eps_n^k$
and find
\begin{equation}
\label{prob3}
p(x,t)\propto\det(P_n(x_i,t))\,\det(\tilde g_k(x_i,t))
\end{equation}
with
\begin{eqnarray}
\nonumber
\tilde g_k(x,t)&=&\sum_{n=0}^\infty
(n+{\textstyle \frac{1}{2}})\,\eps_n^k \,P_n(x)\,e^{-\eps_n t}
=(-1)^k\left(\frac{d}{dt}\right)^k\,g_0(x,t)\\
\label{gkdef2}
&=&\frac{1}{t^k}\left(\frac{d}{ds}\right)^k\,g_0(x,t(1-s))\Big|_{s=0}
\quad.
\end{eqnarray}
The quantity $g_0(x,t)$ can be related to the heat kernel on the
unit sphere $S_2$ with initial condition at
the origin, which yields a useful approximation in the limit
$t\ll 1$
\begin{equation}
\label{gapprox}
g_0(\cos(2\varphi),t)\simeq \frac{1}{8t}\,e^{-\varphi^2/(4t)}
\quad.
\end{equation}
{}From the expansion
\begin{equation}
\label{laguerreexpand}
\frac{1}{1-s}\,e^{-xs/(1-s)} =\sum_{k=0}^\infty \frac{L_k(x)}{k!} s^k
\end{equation}
(here $L_k(x)$ denote the Laguerre polynomials)
and (\ref{gapprox}), (\ref{gkdef2}) we find
\begin{equation}
\label{gksmallt}
\tilde g_k(\cos(2\varphi),t)=\frac{1}{8t^{k+1}}\,
e^{-\varphi^2/(4t)}\,L_k\left(\frac{\varphi^2}{4t}\right)\quad,
\quad t\ll 1\quad.
\end{equation}
This results leads to a probability density (now for the variables
$\varphi_j$ )
\begin{equation}
\label{prob4}
\tilde p(\varphi,t)\simeq C(t)\prod_j\left( \sin(2\varphi_j)\,
e^{-\varphi_j^2/(4t)}\right)\prod_{i<j}\left\{\left(
\sin^2\varphi_i-\sin^2\varphi_j\right)\left(\varphi_i^2-\varphi_j^2
\right)\right\}
\end{equation}
where $C(t)$ is a normalization constant which depends on $t$.
If we write (\ref{prob4}) under the form of a ``Gibbs factor''
$\tilde p(\varphi,t)\sim \exp[-\beta(
\sum_{i<j} u(\varphi_i,\varphi_j)+\sum_i V(\varphi_i,t))]$, we find
that the fictitious corresponding Hamiltonian has the pairwise
interaction and the one body confining potential:
\begin{eqnarray}
\label{interact}
u(\varphi_i,\varphi_j) & = & -\frac{1}{2}\left(\ln|\sin^2\varphi_i-
\sin^2\varphi_j|+\ln|\varphi_i^2-\varphi_j^2|\right)\quad, \\
V(\varphi_i,t) & = & \frac{\varphi_i^2}{8t}-\frac{1}{2}\ln|
\sin(2\varphi_i)|\quad.
\end{eqnarray}

 This can be simplified when $t$ is very small, i.e. $t\ll t_c=1/(8N)$
since $\sin^2\varphi_i\simeq \varphi_i^2$ in this limit. One gets a
joint probability distribution (\ref{prob4}) identical to the so-called
``Laguerre ensemble'' of Ref. \cite{slevin1}, with a logarithmic repulsion
between the $\varphi_i^2$ (see also Ref. \cite{review_matrix}).
For $N\gg 1$, the density reduces to a quarter circle law:
\begin{equation}
\label{semicircle1}
\tilde R_1(\varphi;t)\simeq \frac{1}{4\pi t}
\sqrt{16 t N-\varphi^2}
\end{equation}
if the argument of the square root is not negative (otherwise
$\tilde R_1(\varphi;t)=0$).
Using $\varphi^2\simeq T=\cosh^{-2}\nu$, one gets
\begin{equation}
\label{semicircle2}
\tilde R_1^{(\nu)}(\nu;t)=\frac{2\sinh\nu}{\cosh^3\nu}
\tilde R_1(\cosh^{-1}\nu;t)
\end{equation}
for the variable $\nu\in[0,\infty]$.

 One can note that the exact pairwise interaction for the $\varphi_j$
variables (\ref{interact}) differs for $1/N\ll t\ll 1$ from the
standard RMT interaction given by (\ref{fbeta}) for $\beta=2$.
However, this concerns essentially the quantitative difference between
$\sin^2\varphi$ and $\varphi^2$ whereas for the quasi one-dimensional
disordered wire considered in Ref. \cite{rejaei} this effect was much
more stronger due to the appearance of $\sinh^2(\ldots)$ instead
of $\sin^2(\ldots)$.

In the sections \ref{section:4}, \ref{section:5}, we will return to this point
and compare the results Eqs. (\ref{corr1}),
(\ref{semicircle2}) with the corresponding expressions obtained for
disordered lines, assuming the $M$-Brownian motion ensemble.

\section{$M$--Brownian motion ensemble for the transmission eigenvalues}

\label{section:3}

  The $S$-Brownian motion ensemble was built on the evolution
law $S(t+\delta t)=S(t)\exp (i\delta X)$ where $\delta X$ was an infinitesimal
hermitian random matrix (Eq. (\ref{S_dynamics})). A Fokker-Planck
equation was deduced,
giving the $t$-dependence of the transmission eigenvalues. A
different Fokker-Planck equation is more natural when one
builds a quasi-one dimensional wire, adding in series small
diffusive blocs \cite{dorokhov,mello2,mello3,review_matrix}.
The evolution law of the matrices $u$ and $v$
(``angular part'') and of the matrix $T$ (``radial part'')
defined  in Eq. (\ref{par}) can indeed be deduced from the
{\it multiplicative} transfer matrix $M$, expressed \cite{jalabert2}
in the same parametrization. The law of $M$ describing the addition
in series of the scatterers is now:
\begin{equation}
\label{M_dynamics}
M(t+\delta t)=M(t)M(\delta t)\quad.
\end{equation}
 The Brownian motion time $t$ is in this case related to the
number of scatterers in series (i.~e.: the length $L$ of
the wire). $M$ can be parametrized with the same coordinates
used for $S$ in Eq. \ref{par}, as detailed in Ref. \cite{jalabert1}
for instance. Assuming that the scatterers are isotropically
distributed, the length dependence of the transmission eigenvalues
for the $M$-Brownian motion ensemble is given by a Fokker-planck equation
\cite{dorokhov,mello2,mello3,review_matrix},
which presents interesting analogies and differences with Eq. (\ref{fp1}). The
comparison of those two Brownian motions, assuming two different
evolution laws, will enable us to describe in details the
similarities and the differences between quantum transmission through
two weakly coupled
cavities ($S$-ensemble) and through a disordered wire ($M$-ensemble).
Before this, having used the Beenakker-Rejaei method for solving the
$S$-Brownian motion ensemble, we have been able to improve this method
and we show in return how to complete the solution of the transmission
problem through a disordered wire. Indeed, we explain how to
calculate exactly the length dependence of $n$ order correlation
functions of the transmission eigenvalues, for arbitrary values of $n$.
This allowes us to establish an important equivalence between
the $M$-Brownian motion ensemble and a more microscopic
approach using the non linear $\sigma$-model.

\subsection{Exact calculation of the correlation functions for $\beta=2$}

\label{section:3a}

 We start with the exact result given in Ref. \cite{rejaei}
for the joint probability distribution  of the variables
$\lambda_j=(1-T_j)/T_j=\sinh^2 x_j$ which satisfies
the Fokker-Planck equation of the $M$-Brownian motion ensemble:
\begin{equation}
\label{probmello}
p(\lambda,t)\propto \prod_{i>j} (\lambda_i-\lambda_j)
\,\det(g_m(\lambda_j,t))\,e^{C_N t}\quad,
\end{equation}
with
\begin{eqnarray}
%\label{mellorhodef}
%\rho(\lambda) & = & \prod_{i>j} (\lambda_i-\lambda_j)\quad,\\
\label{mellogndef}
g_m(\lambda,t) & = &
\int_0^\infty dk\,{\textstyle (k^2)^m\ \frac{k}{2}\tanh\left(
\frac{\pi k}{2}\right)}\,P_{\frac{1}{2}(ik-1)}(1+2\lambda)
e^{-k^2 t}\quad,
\end{eqnarray}
where $C_N$ is just given by (\ref{constcn}) for $\beta=2$. We underline
that the changed sign in the exponential factor of (\ref{probmello}) if
compared with (\ref{xgreen0}) is correct.
The variable $t$ is now $t=L/(4Nl)=L/(2\xi)$ ($l$ is the
elastic mean free path,
$L$ is the length of the conductor, and $\xi=2Nl$ is the quasi one-dimensional
localization length for the unitary case).
The index $m$ in (\ref{mellogndef}) takes the values
$m=0,1,\ldots,N-1$. For later use we mention that the generalized
Legendre function can be expressed as a Hypergeometric function via
\begin{equation}
\label{hypergeo}
P_{\frac{1}{2}(ik-1)}(1+2\lambda) = {
\textstyle F\left(\frac{1}{2}-i\frac{k}{2},\frac{1}{2}+i\frac{k}{2};
1;-\lambda\right)}\quad.
\end{equation}
This function is an eigenfunctions of the differential operator
\begin{equation}
\label{diffdef}
D=-\left(4\frac{\partial}{\partial \lambda}\,\lambda(1+\lambda)
\frac{\partial}{\partial \lambda}+1\right)
\end{equation}
with eigenvalue $k^2$. The operator $D$ is of course related with
the one particle Hamiltonian ${\cal H}_0$ used in Ref. \cite{rejaei} by a
suitable transformation. We can now state the following properties of
$g_m(\lambda,t)$
\begin{eqnarray}
\label{gncalc}
g_m(\lambda,t) & = & D^m\,g_0(\lambda,t)\quad,\\
\label{g0calc}
\frac{\partial}{\partial t}\,g_0(\lambda,t) & = & -D\,g_0(\lambda,t)\quad,\\
\label{g0delta}
g_0(\lambda,0) & = & \delta(\lambda)\quad.
\end{eqnarray}
Eqs. (\ref{gncalc}), (\ref{g0calc}) follow directly from (\ref{mellogndef}).
Eq. (\ref{g0delta}) reflects the initial condition for the one particle
Green's function \cite{rejaei} and was already given in \cite{mello1},
where the expression (\ref{probmello}) was found for the one channel case.

We are now interested in integrals of the form
\begin{equation}
\label{coeffdef}
a_{nm}(t)=\int_0^\infty d\lambda\ P_n(1+2\lambda)\,g_m(\lambda,t)
\end{equation}
where $P_n(\cdots)$ again denotes the Legendre polynomial of degree $n$.
We consider first the case $m=0$. From (\ref{g0delta}) we find
$a_{n0}(0)=P_n(1)=1$ and from (\ref{g0calc}) we get the differential
equation
\begin{eqnarray}
\nonumber
\frac{\partial}{\partial t} a_{n0}(t)&=&-\int_0^\infty d\lambda
\ P_n(1+2\lambda)\,D\,g_0(\lambda,t)\\
\label{andiffeq}
&=&-\int_0^\infty d\lambda\ g_0(\lambda,t)\,D\, P_n(1+2\lambda)=
(1+2n)^2\,a_{n0}(t)=\eps_n\,a_{n0}(t)
\end{eqnarray}
which has the solution $a_{n0}(t)=\exp(\eps_n t)$. In Eq. (\ref{andiffeq})
we have used that $P_n(1+2\lambda)$ is an eigenfunction of $D$ with
eigenvalue $-\eps_n=-(1+2n)^2$, a property following directly from
Legendre's differential equation. From Eqs. (\ref{gncalc}),
(\ref{coeffdef}) we get by a similar calculation the general result
\begin{equation}
\label{anresult}
a_{nm}(t)=(-\eps_n)^m\,a_{n0}(t)=(-\eps_n)^m\,e^{\eps_n t}\quad.
\end{equation}
As in Sec. \ref{section:2c} we can replace the exponential
factors $(k^2)^m$ in
Eq. (\ref{mellogndef}) by an arbitrary set of linear independent
Polynomials of (maximal) degree $N-1$ in the variable $k^2$ because the
determinant in Eq. (\ref{probmello}) yields only
a constant factor after such a transformation. We now choose the
replacement
\begin{equation}
\label{melloreplace}
(k^2)^m\quad\to\quad {\textstyle L_m\left(-\frac{1}{4} k^2\right)}
\end{equation}
where $L_m(u)$ are just the Lagrangian interpolation polynomials
already given by Eq. (\ref{lagdef}). We introduce the
following notations
\begin{equation}
\label{melloqndef}
Q_n(\lambda,t)=P_n(1+2\lambda)\,e^{-\eps_n t}\quad,
\end{equation}
and
\begin{equation}
\label{mellohndef}
h_m(\lambda,t)=\int_0^\infty dk\,
{\textstyle L_m\left(-\frac{1}{4} k^2\right)\,
\frac{k}{2}\tanh\left(
\frac{\pi k}{2}\right)}\,P_{\frac{1}{2}(ik-1)}(1+2\lambda)
e^{-k^2 t}\quad.
\end{equation}
{}From Eq. (\ref{anresult}) we get directly for $n,m=0,1,\ldots,N-1$
\begin{equation}
\label{melloorth}
\int_0^\infty d\lambda\ Q_n(\lambda,t)\,h_m(\lambda,t)=
{\textstyle L_m(\frac{1}{4}\eps_n)}=\delta_{nm}\quad.
\end{equation}
The joint probability distribution (\ref{probmello}) can then
be written in the form
\begin{equation}
\label{probmello2}
p(\lambda,t)=\frac{1}{N!}\,\det(Q_n(\lambda_j,t))\,
\det(h_m(\lambda_j,t))=\frac{1}{N!}\,\det
\left(K_N(\lambda_i,\lambda_j;t)_{\,1\le i,j\le N}\right)
\end{equation}
with
\begin{equation}
\label{mellokndef}
K_N(\lambda,\tilde\lambda;t)=\sum_{m=0}^{N-1}
Q_m(\lambda,t)\,h_m(\tilde\lambda,t)\quad.
\end{equation}
Similarly as in Sec. \ref{section:2c} we find from (\ref{melloorth})
\begin{equation}
\label{melloknnorm}
\int_0^\infty d\lambda\ K_N(\lambda,\lambda;t)=N
\end{equation}
and
\begin{equation}
\label{melloknkonv}
\int_0^\infty d\mu\ K_N(\lambda,\mu;t)\,K_N(\mu,\tilde\lambda;t)=
K_N(\lambda,\tilde\lambda;t)\quad.
\end{equation}
These properties lead as in Eq. (\ref{corresult})
to the $m$-point correlation functions
\begin{equation}
\label{mellocorrfunc}
R_m(\lambda_1,\ldots,\lambda_m;t)=\det
\left(K_N(\lambda_i,\lambda_j;t)_{\,1\le i,j\le m}\right)
\end{equation}
where $K_N(\lambda,\tilde\lambda;t)$ can also be expressed by
the more explicit formula
\begin{eqnarray}
\nonumber
K_N(\lambda,\tilde\lambda;t)&=&\int_0^\infty dk\,
{\textstyle \frac{k}{2}\tanh\left(
\frac{\pi k}{2}\right)}\,P_{\frac{1}{2}(ik-1)}(1+2\tilde\lambda)
e^{-k^2 t}\\
\label{melloknexplizit}
&&\times\sum_{m=0}^{N-1} {\textstyle L_m\left(-\frac{1}{4} k^2\right)}\,
P_m(1+2\lambda)\,e^{-\eps_m t}\quad.
\end{eqnarray}

\subsection{Equivalence with the microscopic non linear $\sigma$-model
	approach}

\label{section:3b}

As a first application we calculate an exact expression
for the average conductance, when the length of the wire
increases. This expression is valid from the
disordered conductor towards the Anderson insulator.
Since the conductance is a linear statistic of the $\lambda_i$,
$g=\sum_i(1+\lambda_i)^{-1}$, we have to the evaluate the
corresponding integral over the density:
\begin{equation}
\label{melloavcond}
\langle g\rangle =\int_0^\infty d\lambda\,\frac{1}{1+\lambda}
\sum_{m=0}^{N-1} Q_m(\lambda,t)\,h_m(\lambda,t)\quad.
\end{equation}
Since $Q_m(\lambda,t)$ is a polynomial of degree $m$ in the variable
$1+\lambda$, we can write
\begin{equation}
\label{qmpol}
\frac{1}{1+\lambda}Q_m(\lambda,t)=\frac{1}{1+\lambda}Q_m(-1,t)+
\,r_{m-1}(\lambda)=\frac{(-1)^m}{1+\lambda}\,e^{-\eps_m t}+
r_{m-1}(\lambda)
\end{equation}
where $r_{m-1}(\lambda)$ is a polynomial of degree $m-1$ which does not
contribute in the integral (\ref{melloavcond}) due to the quasi
orthogonality relation (\ref{melloorth}). We find therefore
\begin{equation}
\label{melloavcond2}
\langle g\rangle =\sum_{m=0}^{N-1} \int_0^\infty dk\,
e^{-(\eps_m+k^2) t}\,(-1)^m\,
{\textstyle L_m\left(-\frac{1}{4} k^2\right)\,
\frac{k}{2}\tanh\left(\frac{\pi k}{2}\right)}\cdot I(k)
\end{equation}
where $I(k)$ stands for the integral
\begin{equation}
\label{ikdef}
I(k)=\int_0^\infty \frac{1}{1+\lambda}\,
{\textstyle F\left(\frac{1}{2}-i\frac{k}{2},\frac{1}{2}+i\frac{k}{2};
1;-\lambda\right)}\quad.
\end{equation}
We use the transformation formula 15.3.4 of Ref. \cite{abramowitz}
for the Hypergeometric
function in (\ref{ikdef}) and expand afterwards the latter in its power
series. The integral over $\lambda$ can then be done and gives the sum
\begin{equation}
\label{iksum}
I(k)=\sum_{n=0}^\infty \frac{(-1)^n}{\frac{1}{2}+n-i\frac{k}{2}}
\ { -\frac{1}{2}+i\frac{k}{2} \choose n}
\end{equation}
where ${a \choose n}$ denotes the generalized Binomial
coefficient which is defined for arbitrary complex values of $a$.
The evaluation of the sum (\ref{iksum}) yields
\begin{equation}
\label{ikresult}
I(k)=\int_0^1 du\ (1-u)^{-\frac{1}{2}+i\frac{k}{2}}
\ u^{-\frac{1}{2}-i\frac{k}{2}}=
{\textstyle \Gamma\left(\frac{1}{2}+i\frac{k}{2}\right)
\ \Gamma\left(\frac{1}{2}-i\frac{k}{2}\right)}\quad.
\end{equation}
We now use the expression (\ref{lagdef}) for the
polynominals $L_m(\cdots)$ and
find for the averaged conductance the expression
\begin{equation}
\label{melloavcond3}
\langle g\rangle = 2 \sum_{m=0}^{N-1} \int_0^\infty dk\,
e^{-((1+2m)^2+k^2) t}\ {\textstyle
k\tanh\left(\frac{\pi k}{2}\right)}\ \frac{2m+1}{k^2+(1+2m)^2}
\ a(N,m,k)
\end{equation}
where we have introduced the coefficient
\begin{eqnarray}
\label{anmkdef}
a(N,m,k)&=&\frac{k^2+(1+2m)^2}{4(2m+1)}\,(-1)^m\,
{\textstyle L_m\left(-\frac{1}{4}k^2 \right)\,
\Gamma\left(\frac{1}{2}+i\frac{k}{2}\right)
\ \Gamma\left(\frac{1}{2}-i\frac{k}{2}\right)}\\
\label{anmkresult}
&=& \frac{\Gamma\left(N+\frac{1}{2}+i\frac{k}{2}\right)
\ \Gamma\left(N+\frac{1}{2}-i\frac{k}{2}\right)}{
\Gamma(N-m)\ \Gamma(N+m+1)}\quad.
\end{eqnarray}
The result (\ref{melloavcond3}) is exact for all values of $N$ and
$t=L/(2\xi)$. In the limit $N\to \infty$ the coefficient
$a(N,m,k)$ is just $1$ and the expression (\ref{melloavcond3})
becomes identical with the microscopic result of
Zirnbauer et al \cite{zirnbauer} based on the supersymmetric
non linear sigma model, in the unitary case.

A similar calculation yields the second moment of $g$
\begin{equation}
\label{mellofluct}
\langle g^2\rangle = \frac{1}{2} \sum_{m=0}^{N-1} \int_0^\infty dk\,
e^{-((1+2m)^2+k^2) t}\ {\textstyle
k\tanh\left(\frac{\pi k}{2}\right)}\ (2m+1)
\ a(N,m,k)\quad.
\end{equation}
We omit the details, since this result can also be derived in a more
direct way by the general identity
\begin{equation}
\label{gderiv}
\langle g^2\rangle=-\frac{1}{4}\frac{\partial \langle g\rangle}
{\partial t}
\end{equation}
which follows directly from the original Fokker-Planck equation
\cite{mello4} for the case $\beta=2$.

This gives the proof that the microscopic supersymmetric approach and the
$M$-Brownian motion ensemble of Ref. \cite{mello2,mello4} are equivalent
in the limit $N\to\infty$, for arbitrary values of $L$, at least
as far as the average and the variance of the conductance are concerned.

\subsection{Quasi-$1d$ localized limit.}

\label{section:3c}

In the localized limit $t\gg 1$ the integral in Eq. (\ref{mellohndef})
can be evaluated by a saddle point approximation. In order to do this,
we have to consider the oscillatory behavior of the Legendre function
for large values of the variable $\lambda$. We use now the transformation
formula 15.3.8 of \cite{abramowitz} and express the
$k$-integration measure as
a product of Gamma functions. Instead of Eq. (\ref{mellohndef})
we get then the modified expression
\begin{equation}
\label{hmmodified}
h_m(\lambda,t)=\frac{1}{4\pi}\int_{-\infty}^\infty dk
\ e^{-k^2 t}\ (1+\lambda)^{-\frac{1}{2}-i\frac{k}{2}}
{\textstyle \ H\left(k,\frac{1}{1+\lambda}\right)}
\end{equation}
with the abbreviation
\begin{equation}
\label{hfuncdef}
{\textstyle \ H\left(k,\frac{1}{1+\lambda}\right)}=
\frac{\Gamma(\frac{1}{2}+i\frac{k}{2})^2}{\Gamma(ik)}
\ {\textstyle F\left(\frac{1}{2}+i\frac{k}{2},
\frac{1}{2}+i\frac{k}{2};1+ik;\frac{1}{1+\lambda}\right)
\ L_m\left(-\frac{1}{4}k^2\right)\quad.}
\end{equation}
It is more convenient to replace $\lambda$ by the new variable
$z=\frac{1}{2}\ln(1+\lambda)$. A simple substitution
then leads to the expression
\begin{equation}
\label{hfuncsub}
h_m(e^{2z}-1,t)=e^{-z^2/(4t)-z}\ \frac{1}{4\pi\sqrt{t}}
\int_{-\infty}^\infty dq\ e^{-q^2}\
H\left(\frac{q}{\sqrt{t}}-i\frac{z}{2t},e^{-2z}\right)
\end{equation}
which is still exact. We mention that this expression is rather
useful for an efficient numerical evaluation
since the difficult oscillatory behavior of
the Legendre functions has been taken into account by a shift of the
integration path into the complex plane. In the limits $t\gg 1$,
$z\gg 1$, the integral
can be done and we obtain up to corrections of the order $1/t$ or
$e^{-2z}$ respectively
\begin{equation}
\label{hmapprox}
h_m(e^{2z}-1,t)\simeq\frac{1}{2\sqrt{4\pi t}}e^{-z^2/(4t)-z}
\ \frac{\Gamma(\frac{1}{2}(1+\frac{z}{2t}))^2}{\Gamma(
\frac{z}{2t})}\ L_m\left(\frac{1}{4}\Bigl(
\frac{z}{2t}\Bigr)^2\right)\quad.
\end{equation}
The corresponding limit for the $Q_m(\lambda,t)$ in Eq. (\ref{melloqndef})
is
\begin{equation}
\label{qmapprox}
Q_m(e^{2z}-1,t)\simeq \frac{(2m)!}{m!^2}\ e^{-(1+2m)^2t}
\ e^{2zm}\quad.
\end{equation}
The density $\tilde R_1(z;t)$ and the two point function
$\tilde R_2(z,\tilde z;t)$
for the variable $z$ can be obtained
from Eq. (\ref{mellocorrfunc}) and Eq. (\ref{mellokndef}). The result
(containing additional factors $2e^{2z}$ or $2e^{2\tilde z}$
due to the jacobian of the variable transformation $\lambda\to z$)
is
\begin{equation}
\label{rz1result}
\tilde R_1(z;t)\simeq \sum_{m=0}^{N-1} G_m(z;t)\quad,\quad
\tilde R_2(z,\tilde z;t)\simeq \sum_{m,\tilde m=0, m\neq \tilde m}^{N-1}
G_m(z;t)\,G_{\tilde m}(\tilde z;t)
\end{equation}
where we have used the abbreviation
\begin{equation}
\label{gaussabri}
G_m(z;t)=\frac{1}{\sqrt{4\pi t}}
{\textstyle \ \exp\left(-\frac{1}{4t}(z-2t(1+2m))^2\right)}
\end{equation}
for the Gaussian distribution. We mention that
for (\ref{rz1result}) we have replaced
the argument $\frac{z}{2t}$ of the Gamma functions and of the polynomial
$L_m(\cdots)$ in Eq. (\ref{hmapprox}) by the
most probable values $\frac{z}{2t}= 1+2m$ or $1+2\tilde m$ due to
the Gaussian factors. We recover the simple sum of Gaussian distributions
with mean values $2t(1+2m)$ and a variance $2t$ already
obtained by a direct simplification of the Fokker Planck equation
valid in the localized limit\cite{pichardzanon,review_matrix,lesarcs}.
The above calculation shows that the general
expressions (\ref{mellocorrfunc}),
(\ref{melloknexplizit}) of Sec. \ref{section:3a} are indeed consistent
with known results concerning the quasi-$1d$ insulators.

\section{Weakly coupled chaotic cavities versus disordered lines}

\label{section:4}

We describe now the differences between transmission through two
ballistic cavities coupled by a narrow constriction
(section \ref{section:2add}) and transmission through an homogeneously
disordered wire of constant transverse section (section \ref{section:3}).
This can be of practical interest for many purposes which are not
restricted only to mesoscopic electrical conductances, but which
could be relevant for other systems as wave guide communication lines.
One can imagine that the signal from an antenna to a receiver is
anomalously weak and that the radio-engineering problem is to know
if the weakness of the signal is due to an homogeneous deterioration
of the transmission line or only to a local deterioration resulting
from an accidental narrow constriction.

 Characteristic behaviors of the $S$-Brownian motion ensemble
are first given, characterizing the coupled ballistic dots under
a small applied magnetic field. In Fig. 2, the density  of
the variable $\nu=(\mbox{arsinh}(\lambda))^{1/2}$
is shown for the one channel case ($N=1$) at four different values
of $t/t_c=10.0,\ 1.0\ ,0.1\ ,0.01$ ($t_c=1/(8N)$ is the typical
time scale introduced in Section \ref{section:2b}). The variables
$\lambda=(1-T)/T=(1+x)/(1-x)$ are the radial parameters of the
transfer matrix, which are usual in random transfer matrix
theory \cite{jalabert1,mello2,pichardzanon}.
The curves are calculated from Eqs. (\ref{corr1}),(\ref{knexp}).
The figure 3 gives this density for the same values of $t$ but
for $N=5$ channels. In addition for $t/t_c=0.1,\ 0.01$ the asymptotic
expression (\ref{semicircle2}) that arises from the Laguerre
approximation is shown, too. The latter is valid if $N\gg 1$ and
$t\ll t_c$. In both figures one can see that the typical values
for $\nu$ increase (on a logarithmic scale) if $t$ decrease.
One can identify a rather sharp minimum value $\nu_{\min}$ for
$\nu$ if $t\ll t_c$. A simple estimation from Eq. (\ref{semicircle1})
yields
\begin{equation}
\label{numin}
\nu_{\min}\simeq -\frac{1}{2}\ln\left(\frac{t}{2t_c}\right)\quad.
\end{equation}
For values $\nu\ge\nu_{\min}$ the distribution is rather broad and
the overall form is (nearly) independent of $N$ (apart from the case
$t\gg t_c$). A reduction of the time leads essentially to
a translation of the density.
For finite values of $N$ (see  $N=5$), one can see typical but rather
small oscillations around the limiting distribution for $N\to\infty$.
The value $t/t_c=10.0$ corresponds within the accuracy of the plots
to the stationary limit $t/t_c\to \infty$, i.e. the single CUE behavior.
For this case the position of the first maximum decreases and becomes
sharper when $N$ increases. This gives rise to a conductance $g\sim N/2$
(cp. Eq. (\ref{gaverage})).

 The corresponding behaviors of the $M$-Brownian motion
ensemble exhibit essential differences from the previous case.
Fig. 4 shows the $\nu$-density for the disordered wire with $N=5$
channels in three different regimes, i.e. the localized
regime ($L/\xi=10.0$), the crossover regime ($L/\xi=1.0$) and the
metallic regime ($L/\xi=0.2$). Figs. 5a-c contain the same density
curves in a more readable scale than in Fig. 4. The gaussian
approximation (\ref{rz1result}) (dotted line) which fits very well
the exact density at $L/\xi=10.0$ is indicated in Fig. 5c.
\footnote{The variable $z$ used in Section \ref{section:3c} is
only approximately equal to $\nu$ ( i.e. $z=\ln(\cosh\nu)\simeq\nu$
when $\nu\gg 1$)}.
The results shown in the Figs. 4, 5 were obtained by a numerical
evaluation of the integral (\ref{hfuncsub}) except for the limits $t\ll 1$,
$\nu\le 1$ where the integral (\ref{mellohndef}) is better suited.
The substitution $z=\ln(\cosh\nu)$ was of course exactly considered.

  One can see that the $\nu$-density of the disordered line differs
essentially from the density shown in Fig. 3 for the two weakly
coupled cavities.  For small transmission ( i.e. the localized limit),
the $\nu$-density is mainly concentrated around maximal values
$\nu_{\max,m}=(1+2m)L/\xi$ ($m=0,1,\ldots,N-1$) with a
relative variance $\langle(\nu-\nu_{\max,m})^2\rangle^{1/2}/\nu_{\max,m}$
\hbox{$\sim\sqrt{\xi/L}$}. Between the maxima, the density nearly
vanishes. This effect can be seen in Fig. 5c and is still
increased if the ratio $L/\xi$ is increased. But in the crossover
regime $L\approx\xi$ (Fig. 5b), the minima are no longer negligible
compared to the maxima. In Fig. 5a ( metallic regime), the amplitude of the
finite $N$ oscillations are small but still clearly observable.
Further calculations for different values of $N$ show
that the amplitude of the oscillations remains unchanged if $N$ is
increased for a fixed ratio $L/\xi$. In order to reach a metallic
regime where the density is really a smooth function of $\nu$,
one has to increase $N$ to very high values and at the same time to
decrease $L/\xi$ (within the limit $L/\xi\gg 1/N$).

Finally, we compare $\langle g\rangle$ and of
$\langle \delta g^2\rangle$ between the two cases in the limit
of small transmission, i.e. $t\ll t_c$ for the $2\times$CUE $\to$
CUE crossover or $L\gg \xi$ for the disordered line.
For the two weakly coupled cavities, we get in the lowest
order of $t/t_c$ from Eqs. (\ref{gaverage}), (\ref{gautocorr})
\begin{eqnarray}
\label{gsmallt}
\langle g(t)\rangle&\simeq & \frac{N}{2}\,\frac{t}{t_c}\simeq n\quad,\\
\label{deltagsmallt}
\langle \delta g^2(t)\rangle &\simeq &\frac{1}{4}\,\left(\frac{t}{t_c}
\right)^2\simeq \left(\frac{n}{N}\right)^2
\end{eqnarray}
where $n\simeq (N/2)\cdot(t/t_c)$ is the effective number of
coupling channels introduced in Section \ref{section:1}.
These results lead to the limit
\begin{equation}
\label{glimit1}
\lim_{t\to 0}\,\frac{\langle \delta g^2(t)\rangle}{\langle g(t)\rangle^2}
=\frac{1}{N^2}
\end{equation}
for the relative variance of $g$. The corresponding expressions
for the quasi one-dimensional line in the localized limit
$L\gg \xi$ can be obtained from Eqs. (\ref{melloavcond3}), (\ref{gderiv})
by an evaluation of the $k$-integral in a saddle point approximation
\begin{equation}
\label{gslok}
\langle g(L)\rangle\simeq \frac{\pi^{3/2}}{4}\,
\frac{\Gamma(N+\frac{1}{2})^2}{\Gamma(N)\Gamma(N+1)}\,
\left(\frac{L}{2\xi}\right)^{-3/2}
\ e^{-L/(2\xi)}\quad,
\end{equation}
\begin{equation}
\label{deltaglok}
\langle \delta g^2(L)\rangle\simeq\langle g^2(L)\rangle\simeq
\frac{1}{4}\langle g(L)\rangle\quad.
\end{equation}
The relative variance of $g$ is now infinite and the infinite
length limit is characterized by
\begin{equation}
\label{glimit2}
\lim_{L\to\infty}\,\frac{\langle \delta g^2(L)\rangle}{\langle g(L)\rangle}
=\frac{1}{4}\quad.
\end{equation}
The comparison with (\ref{glimit1}) shows again a
qualitative difference in the statistics of the conductance
between the two cases. Eqs. (\ref{gslok}) and (\ref{deltaglok})
illustrate the well known fact that the average of the conductance
does not significantly characterize the conductance properties
in the localized limit. On the other hand, the limit (\ref{glimit1})
for the two weakly coupled ballistic cavities is a small quantity
for a large number of channels.

\section{Summary}

\label{section:5}

 The $S$ and $M$ Brownian motion ensembles which we have used are
characterized by two Fokker-Planck equations expressing isotropic
diffusion equations on the compact space of unitary matrices ($S$)
or on a non compact symmetric space of pseudo-unitary transfer
matrices \cite{hueffmann} respectively. The differential
operators which appear in the Fokker-Planck equations are just the
radial part of the Laplace-Beltrami operators of the corresponding
spaces, describing free isotropic diffusion on those curved spaces.
In the case of the $M$-Brownian motion ensemble, the variable $\nu$ is
the good coordinate since it measures in some way the geodesic distance
on the space of pseudo-unitary matrices (cp. Ref. \cite{hueffmann}).
Fig. 4 shows very clearly the free diffusion of $\nu$. Since the range
of $\nu$ is non compact the rather strong level repulsion gives
well separated maxima. For the compact space of unitary matrices,
it is the variable $\varphi$ used in the Sections \ref{section:2a}
and \ref{section:2b} which measures the geodesic distance. The range
for $\varphi$ is compact and diffusion leads after a long time to an
essentially uniform distribution. The level repulsion only causes some
small oscillations around this distribution.

Apart from the (important) differences which we have reviewed, we
can also notice nice mathematical similarities between the two
Fokker-Planck approaches. Many formulas and results of
Section \ref{section:2} coincide with the results of Ref. \cite{rejaei}
and Section \ref{section:3}, after replacing the variables $\varphi_j$
by the $\nu_j$ ($=x_j$ in \cite{rejaei}) and $\sin,\,\cos$ by
$\sinh,\,\cosh$. This is partly related to the similarity of the
parametrization (\ref{par}) with the corresponding parametrization
of the transfer matrix\cite{jalabert2} (with the substitution
$T \to -\Lambda$). This occurs only for the unitary case ($\beta=2$).
For the orthogonal and unitary cases ($\beta=1,4$) this
similarity does not exist, since the unitary matrices $u_{1,2}$, $v_{1,2}$
in (\ref{par}) are related in a different way than for the transfer matrix
\cite{jalabert2} by time reversal symmetry.

 Our results (Eqs. (\ref{gsmallt})-(\ref{glimit2}) are valid for
an arbitrary number $N=1,2,3,\ldots$ of channels. In the case of a
single channel, the quotient of the Gamma functions takes the
value $\pi/4$ and Eq. (\ref{gslok}) becomes identical with
the result of Ref. \cite{frahm2} obtained using a supervector
model for the one-dimensional white noise potential at small disorder.
Indeed, it was known that the one-dimensional white noise model
for small disorder leads \cite{kree1,kumar1} to the same
Fokker-Planck equation as the one of Refs.
\cite{dorokhov,mello1,mello2} for one channel.
In the opposite limit $N\to\infty$, the Gamma function quotient
reduces to one and we recover the corresponding expression of
Ref. \cite{zirnbauer} obtained by the supersymmetric non linear
$\sigma$-model:
the expressions (\ref{melloavcond3}), (\ref{mellofluct})
for $\langle g\rangle$ and $\langle g^2\rangle$ coinciding exactly
for all length scales $0<L/\xi<\infty$ with those of Ref. \cite{zirnbauer}.
The non linear sigma model for quasi one-dimensional
disordered conductors can at least be derived assuming three different
microscopic Hamiltonians, i.e. the white noise potential \cite{efetov},
Wegner's $N$ orbital model  \cite{iida1}, and a random banded matrix
Hamiltonian \cite{fyodorov}. It is then very remarkable that in the
unitary case, these microscopic approaches become equivalent to the
$M$-Brownian motion ensemble of Ref. \cite{mello2}.
This behavior is apparently not true for the symplectic case, since
the non-linear sigma model predicts \cite{zirnbauer} finite values of
$\langle g\rangle$ and $\langle \delta g^2\rangle$ in the localized regime,
while the Fokker-Planck equation gives an exponential decrease of the
typical conductance for each $\beta=1,2,4$\cite{review_matrix}. In
principle, it could be still possible that in the large
$N$ limit the average $\lim_{L\to\infty} \langle g(L)\rangle$ takes a
non vanishing value because of some very subtle effects concerning
the tails of the normal distribution of $\ln(g)$. But even this cannot
happen, as it can be easily seen from the generalization \cite{mello4}
of Eq. (\ref{gderiv}) for arbitrary $\beta$
\begin{equation}
\label{gderiv_beta}
\frac{\partial \langle g\rangle}{\partial L}=
-\frac{2}{\xi_\beta}\, \left(
\langle g_2\rangle + \frac{\beta}{2}\left(
\langle g^2\rangle-\langle g_2\rangle\right)\right)\quad.
\end{equation}
Here $\xi_\beta\simeq \beta \xi_1$ denotes the localization length and
$g_2$ is just the trace of the squared transmission matrix $(tt^\dagger)^2$,
i.e. $g_2=\sum_i (1+\lambda_i)^{-2}$. From the inequality
$0\le g_2 \le g^2$ one finds directly that in the limit $L\to\infty$
the behavior $\langle g\rangle =const. \neq 0$ (as predicted in
Ref. \cite{zirnbauer} for $\beta=4$) is not possible since
$\langle g^2\rangle$ must vanish, which yields directly
$\langle g\rangle =0$.
This contradiction when $\beta=4$ between the two approaches is surely
worth to be considered in future work.

 Though the two Fokker-Planck approaches are very similar
from a mathematical point of view, they nevertheless describe two
different physical situations where the Fokker-Planck time controls
the crossover between systems with a high and a low conductance.
 For both approaches we have found (rather) closed
expressions (valid for a finite channel number) for
any $m$-point correlation functions of the transmission eigenvalues,
allowing us to calculate the first and second moment of the
conductance. This complete analytical solution
is essentially restricted to the unitary case where both
Fokker Planck approaches can be solved in terms of fermionic one particle
Green's functions. At the moment,  the extension of this solution
for arbitrary $N$ to the orthogonal and symplectic cases looks difficult,
since the Hamilton operator obtained after the Sutherland transformation
does not reduces to the sum of one particle operators.

\section*{Acknowledgment}

\label{Acknowledgments}

 We thank B. L. Altshuler and C. W. J. Beenakker for helpful
discussions. In particular, our understanding of the universality
of the transmission eigenvalue correlation functions has been
improved thank to stimulating remarks by C. W. J. Beenakker.
This work was supported in part by EEC, contract No. SCC-CT90-0020.
Klaus Frahm acknowledges the D.F.G. for a post-doctoral fellowship.

\appendix

\section*{Fokker-Planck equation for the transmission eigenvalues
in the $S$-Brownian motion ensemble.}

\label{append:a}

In this appendix, we derive the Fokker-Planck equation (\ref{fp1})
for the case $\beta=2$. In order to do this, we consider the matrix
\begin{equation}
Q(t)=S(t)\Sigma_z S(t)^\dagger \Sigma_z=
\left(\begin{array}{cc}
v_1 & 0 \\
0 & u_1 \\
\end{array}\right)
\left(\begin{array}{cc}
\cos(2\varphi) & i\sin(2\varphi) \\
i\sin(2\varphi) &\cos(2\varphi) \\
\end{array}\right)
\left(\begin{array}{cc}
v_1^\dagger & 0 \\
0 & u_1^\dagger \\
\end{array}\right)
\end{equation}
where
\begin{equation}
\label{sigz_def}
\Sigma_z=
\left(\begin{array}{cc}
1 & 0 \\
0 & -1 \\
\end{array}\right)\quad.
\end{equation}
The matrix $Q(t+\delta t)$ then has the form
\begin{eqnarray}
\nonumber
Q(t+\delta t)&=&
\left(\begin{array}{cc}
v_1(t+\delta t) & 0 \\
0 & u_1(t+\delta_t) \\
\end{array}\right)\\
\label{pert0}
&&\times\left(\begin{array}{cc}
\cos(2(\varphi+\delta\varphi)) & i\sin(2(\varphi+\delta\varphi)) \\
i\sin(2(\varphi+\delta\varphi)) &\cos(2(\varphi+\delta\varphi)) \\
\end{array}\right)
\left(\begin{array}{cc}
v_1^\dagger(t+\delta t) & 0 \\
0 & u_1^\dagger(t+\delta_t) \\
\end{array}\right)\quad.
\end{eqnarray}
On the other hand the evolution equation $S(t+\delta t)=S(t) e^{i\delta X}$
implies
\begin{equation}
\label{pert1}
Q(t+\delta t)=S(t) e^{i\delta X} e^{-i\Sigma_z\delta X\Sigma_z}
\Sigma_z S(t)^\dagger\Sigma_z=S(t)e^{i\delta Q}\Sigma_z S(t)^\dagger\Sigma_z
\end{equation}
with
\begin{equation}
i\delta Q=i(\delta X-\Sigma_z\delta X\Sigma_z)+\frac{1}{2}\left[
\delta X,\Sigma_z\delta X\Sigma_z\right]+{\cal O}(\delta X^3)\quad.
\end{equation}
The matrix $\delta Q$ has the form
\begin{equation}
\delta Q=
\left(\begin{array}{cc}
0 & \delta m \\
\delta m^\dagger & 0 \\
\end{array}\right)
\end{equation}
where $\delta m$ is an arbitrary complex $N\times N$-random matrix with
\begin{equation}
\label{stat}
\langle \delta m_{ij}\rangle =0\quad,\quad
\langle \delta \bar m_{ij}\delta m_{kl}\rangle=4D \delta t
\ \delta_{ik}\delta_{jl}\quad,\quad
\langle \delta m_{ij}\delta m_{kl}\rangle=0\quad.
\end{equation}
The matrix $S(t)$ in (\ref{pert1}) contains still the unitary matrices
$v_2$ and $u_2$ which can be taken into account by the replacement
$v_2\delta m u_2^\dagger=\delta \tilde m$, This gives for
$\delta \tilde m$ the same statistics than those described by
Eq. (\ref{stat}).

At this point, we can make an important remark concerning the case
where the matrix $\delta X$ obeys a more general distribution than
that of Eq. (\ref{x_statistic}). In this case, the property
(\ref{stat}) for the matrix $\delta m$ is of course no longer valid.
But let us assume that the two unitary matrices $u_2$ and $v_2$ are
independently distributed according to the invariant measure for
unitary $N\times N$-matrices, i.e. $u_2$ and $v_2$ are described
by two independent $N\times N$ CUEs. Then the average over these two
unitary matrices yields just the distribution (\ref{stat}) for the
transformed matrix $\delta\tilde m=v_2\delta m u_2^\dagger$. The
only information that we need from $\delta m$ is therefore the invariant
quantity
\begin{equation}
\label{dm_inform}
\langle\mbox{tr}(\delta m^\dagger\delta m)\rangle=4D\,N^2\ \delta t
\end{equation}
which relates the time step $\delta t$ with the perturbation $\delta X$
or $\delta m$ respectively. This proves that the Fokker-Planck equation
derived below is also valid for a more general perturbation $\delta X$,
provided that the two unitary matrices $u_2$ and $v_2$ are for each time
$t$ independently CUE distributed.
The latter assumption is of course of crucial importance. In the
case of an initial condition of one $2N\times 2N$ COE or two
$N\times N$ CUEs for $S(0)$, this assumption is at least valid
at $t=0$ and we assume that it remains correct for arbitrary $t>0$.
We conclude that the Fokker-Planck equation derived for a perturbation
$\delta X$ obeying (\ref{x_statistic}), still holds if one can relate
the Fokker-Planck time to the perturbation via (\ref{dm_inform}).

Since we are
interested in the statistics of the $\delta\varphi_j$ in (\ref{pert0}),
we have to express the $\delta \varphi_j$ in terms of $\delta \tilde m$.
This corresponds to an eigenvalue problem after the transformation
\begin{equation}
\label{uu_def}
\tilde Q=U^\dagger Q U\quad,\quad U=\frac{1}{\sqrt 2}
\left(\begin{array}{cc}
1 & 1 \\
1 & -1 \\
\end{array}\right)
\end{equation}
which implies
\begin{eqnarray}
\label{a10}
\tilde Q(t)&=&
W\cdot\left(\begin{array}{cc}
e^{2i\varphi} & 0 \\
0 & e^{-2i\varphi} \\
\end{array}\right)\cdot W^\dagger\quad,\\
\label{a11}
\tilde Q(t+\delta t)&=&
W\cdot\left(\begin{array}{cc}
e^{i\varphi} & 0 \\
0 & e^{-i\varphi} \\
\end{array}\right)
e^{i\delta\tilde Q}
\left(\begin{array}{cc}
e^{i\varphi} & 0 \\
0 & e^{-i\varphi} \\
\end{array}\right)\cdot W^\dagger\quad,\\
\label{dmdef}
\delta\tilde Q &=&\frac{1}{2}
\left(\begin{array}{cc}
\delta \tilde m^\dagger+\delta \tilde m &
\delta \tilde m^\dagger-\delta \tilde m \\
-(\delta \tilde m^\dagger-\delta \tilde m) &
-(\delta \tilde m^\dagger+\delta \tilde m) \\
\end{array}\right)\quad.
\end{eqnarray}
The matrix $W$ which appears in (\ref{a10}) and (\ref{a11}) is
just the unitary matrix
\begin{equation}
\label{ww_def}
W=U^\dagger\,
\left(\begin{array}{cc}
v_1 & 0 \\
0 & u_1 \\
\end{array}\right)\,U
\end{equation}
where $U$ is defined in (\ref{uu_def}). The matrix $W$ results only
in a similarity transformation which does not change the eigenvalues.
Eq. (\ref{pert0}) implies that $\tilde Q(t+\delta t)$ has just the
eigenvalues $e^{2i(\varphi_j+\delta\varphi_j)}$ and
$e^{-2i(\varphi_j+\delta\varphi_j)}$, which can be obtained by second order
perturbation theory
\begin{equation}
\label{pert2}
e^{2i(\varphi_j+\delta\varphi_j)}=e^{2i\varphi_j}+V_{jj}+
\sum_{k(\neq j)} \frac{V_{jk} V_{kj}}{e^{2i\varphi_j}-e^{2i\varphi_k}}+
\sum_{k} \frac{V_{j,k+N} V_{k+N,j}}{e^{2i\varphi_j}-e^{-2i\varphi_k}}
\quad.
\end{equation}
The perturbing matrix $V$ in Eq. (\ref{pert2}) is just
\begin{equation}
V=
\left(\begin{array}{cc}
e^{i\varphi} & 0 \\
0 & e^{-i\varphi} \\
\end{array}\right)
\left(i\delta\tilde Q-
\frac{1}{2}\delta\tilde Q^2\right)
\left(\begin{array}{cc}
e^{i\varphi} & 0 \\
0 & e^{-i\varphi} \\
\end{array}\right)\quad.
\end{equation}
Eq. (\ref{pert2}) leads after some algebra to
\begin{equation}
\label{pert3}
\delta \varphi_j=\frac{1}{2}\delta\tilde Q_{jj}+
\frac{1}{4}\biggl(\sum_{k(\neq j)} \cot(\varphi_j-\varphi_k)
\delta\tilde Q_{jk}\,\delta\tilde Q_{kj}
+\sum_{k} \cot(\varphi_j+\varphi_k)
\delta\tilde Q_{j,k+N}\,\delta\tilde Q_{N+k,j}\biggr)\quad.
\end{equation}
We insert Eq. (\ref{dmdef}) in Eq. (\ref{pert3}) and apply the average
(\ref{stat}) for the matrix $\delta \tilde m$ that results in
\begin{equation}
\langle \delta\varphi_j \rangle =\frac{D}{4} f_j(\varphi)\ \delta t\quad,
\quad \langle\delta\varphi_j\delta\varphi_k\rangle =\frac{D}{2}
\ \delta t\ \delta_{jk}
\end{equation}
with
\begin{equation}
f_j(\varphi)=2\sum_{k(\neq j)}\Bigl(\cot(\varphi_j-\varphi_k)+
\cot(\varphi_j+\varphi_k)\Bigr)+2\cot(2\varphi_j)\quad.
\end{equation}
The corresponding Fokker-Planck equation has the form
\begin{equation}
\label{fp2}
\frac{\partial p(\varphi,t)}{\partial t}=
\frac{D}{4}\sum_j \frac{\partial}{\partial \varphi_j}
\left(\left(\frac{\partial}{\partial \varphi_j}-f_j(\varphi)\right)
p(\varphi,t)\right)
\end{equation}
which is just Eq. (\ref{fp1}) for $\beta=2$ because $f_j(\varphi)=
\frac{\partial}{\partial \varphi_j}\ln F_2(\varphi)$ with
$F_2(\varphi)$ given by Eq. (\ref{fbeta}).
%\end{document}

\vfill\eject
\centerline{\bf FIGURE CAPTIONS}

\begin{enumerate}

\item[Fig. 1\phantom{a}]
	Two chaotic ballistic cavities coupled with a $n$ channel
        contact and each of them connected to an electron reservoir
	through a $N$ channel contact.

\item[Fig. 2\phantom{a}]
	Density of the variable $\nu=(\mbox{arsinh}(\lambda))^{1/2}$
	for the crossover $2\times$CUE $\to$ CUE with $N=1$ at
	different values of $t/t_c=10.0,\ 1.0,\ 0.1,\ 0.01$.
	The curves are calculated from
	Eqs. (\ref{corr1}) and (\ref{knexp}).

\item[Fig. 3\phantom{a}]
	Density of the variable $\nu=(\mbox{arsinh}(\lambda))^{1/2}$
	for the crossover $2\times$CUE $\to$ CUE with $N=5$ at
	different values of $t/t_c=10.0,\ 1.0,\ 0.1,\ 0.01$.
	The curves are calculated from
	Eqs. (\ref{corr1}) and (\ref{knexp}).
	In addition, for $t/t_c=0.1,\ 0.01$
	the Laguerre approximation (\ref{semicircle2}) is
	included.

\item[Fig. 4\phantom{a}]
	Density of the variable $\nu=(\mbox{arsinh}(\lambda))^{1/2}$
	for the disordered line at three different
	length scales, $L/\xi=0.2,\ 1.0,\ 10.0$. The curves are
	obtained from a numerical evaluation of Eq. (\ref{hfuncsub}).

\item[Fig. 5a]
	Same as Fig. 4 for $L/\xi=0.2$ but with a modified scale.

\item[Fig. 5b]
	Same as Fig. 4 for $L/\xi=1.0$ but with a modified scale.

\item[Fig. 5c]
	Same as Fig. 4 for $L/\xi=10.0$ but with a modified scale.
	The Gaussian approximation (Eq. (\ref{rz1result}), dotted line)
        is included for comparison.

\end{enumerate}

\vfill\eject
\includegraphics{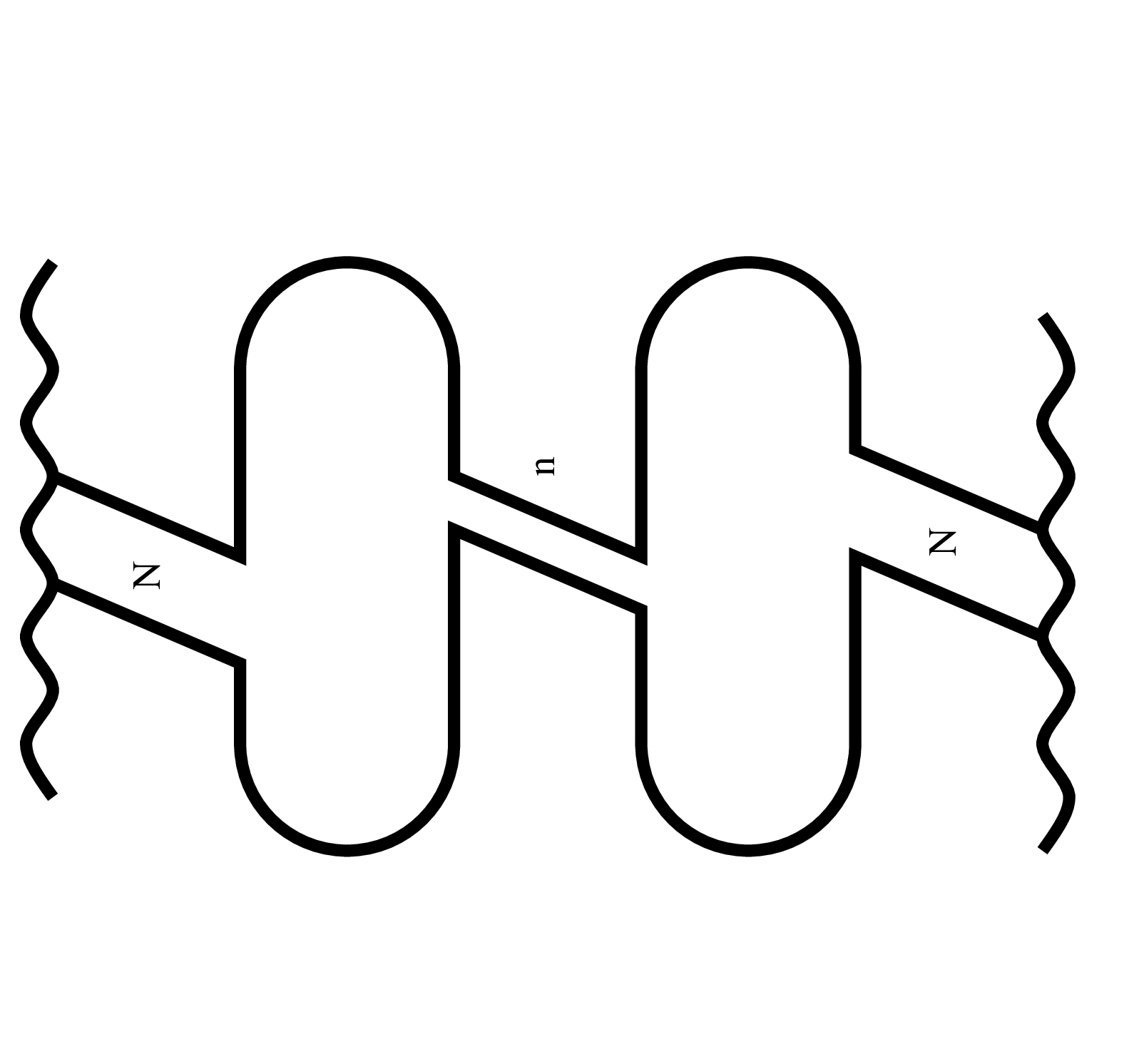}
%\phantom{aaaa}
\centerline{\LARGE Fig. 1}
\vfill\eject
\includegraphics{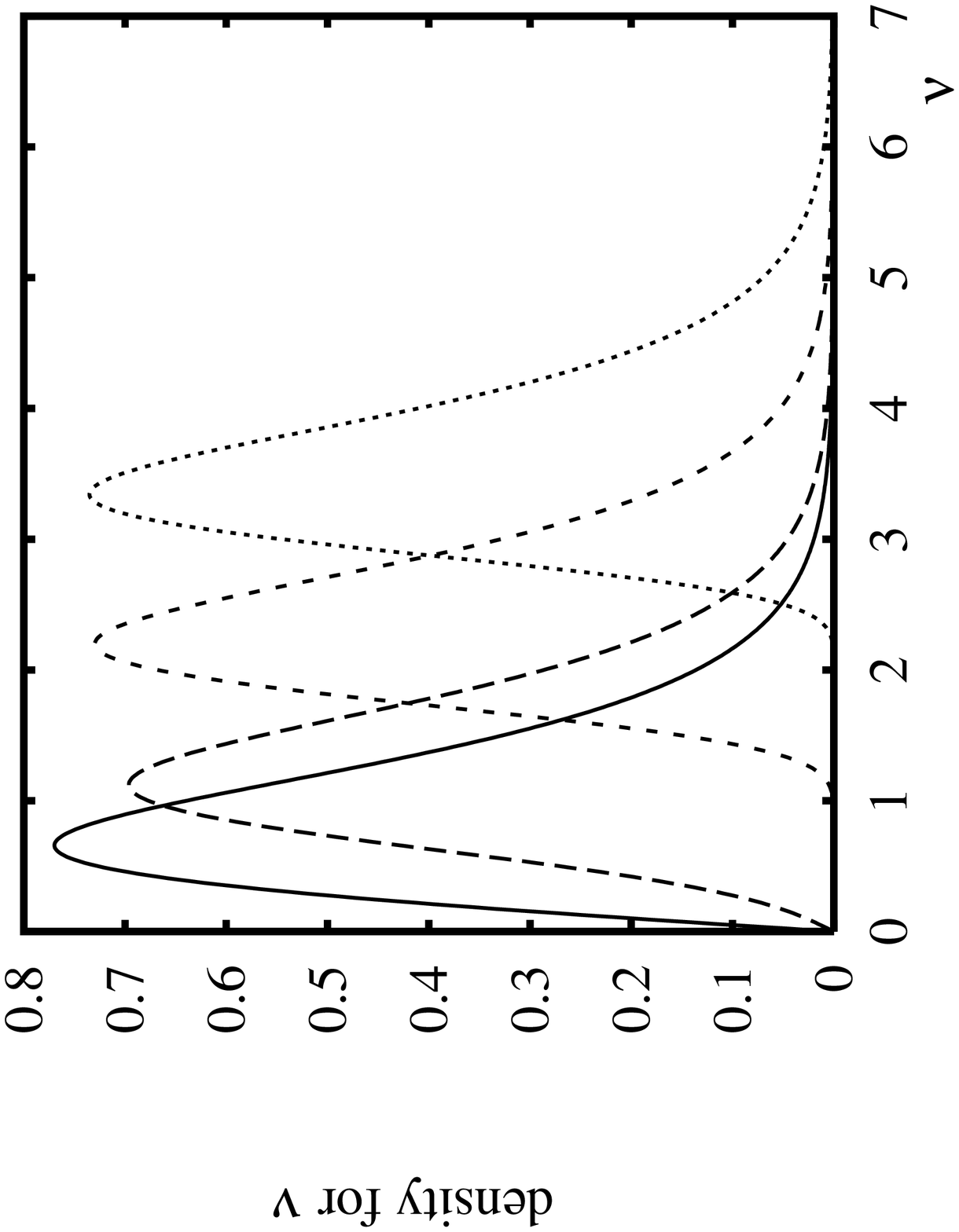}
%\phantom{aaaa}
\centerline{\LARGE Fig. 2}
\vfill\eject
\includegraphics{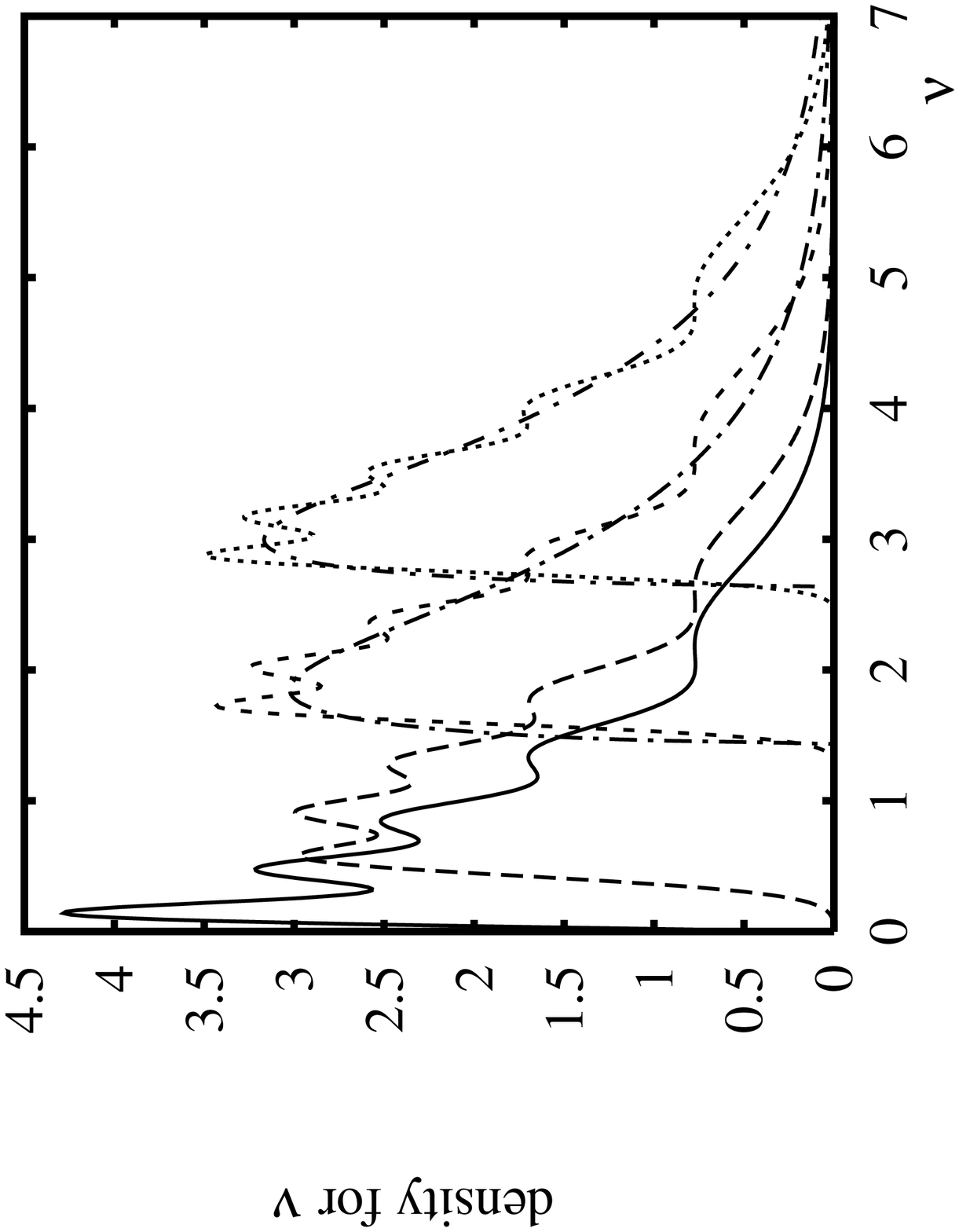}
%\phantom{aaaa}
\centerline{\LARGE Fig. 3}
\vfill\eject
\includegraphics{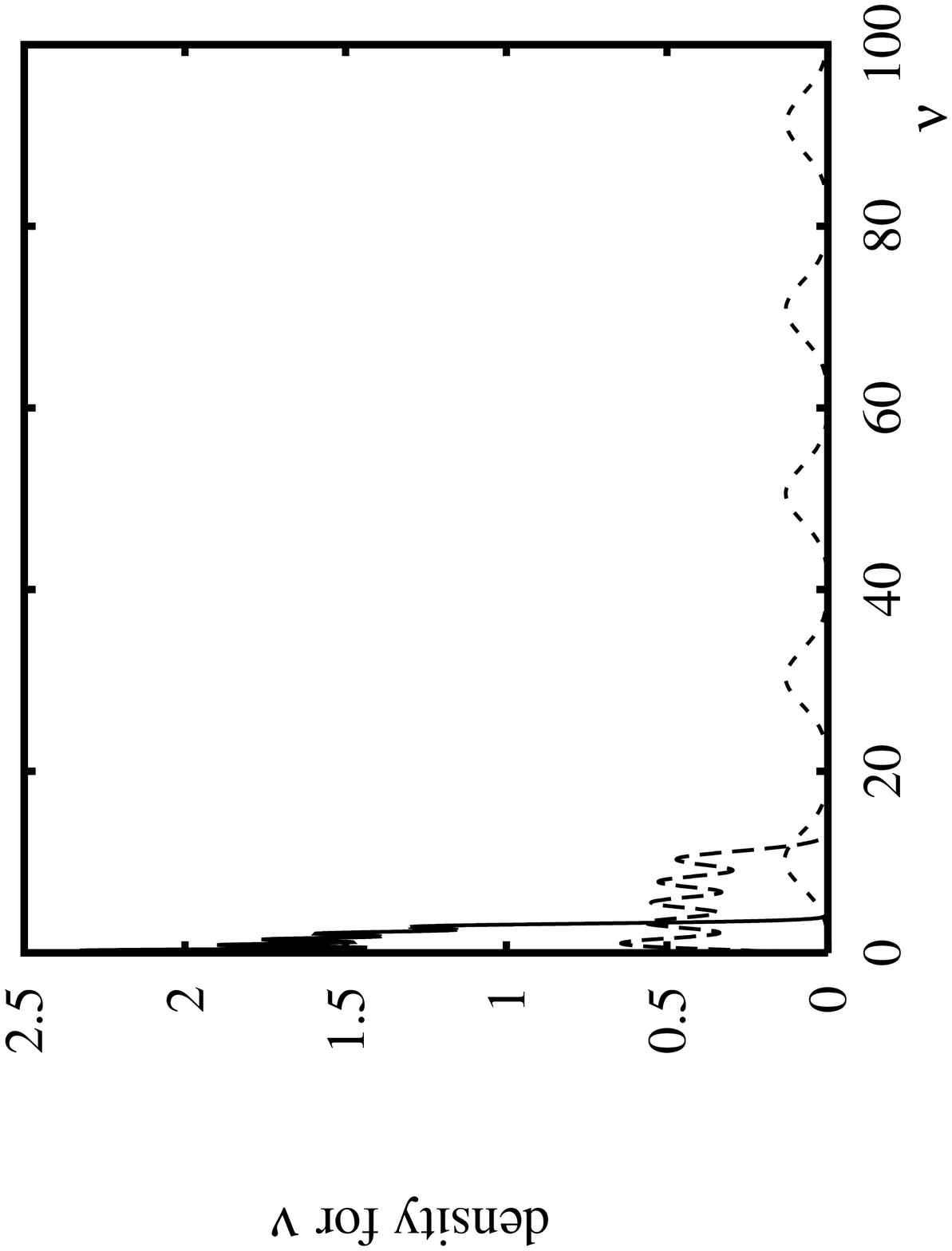}
\centerline{\LARGE Fig. 4}
%\phantom{aaaa}
\vfill\eject
\includegraphics{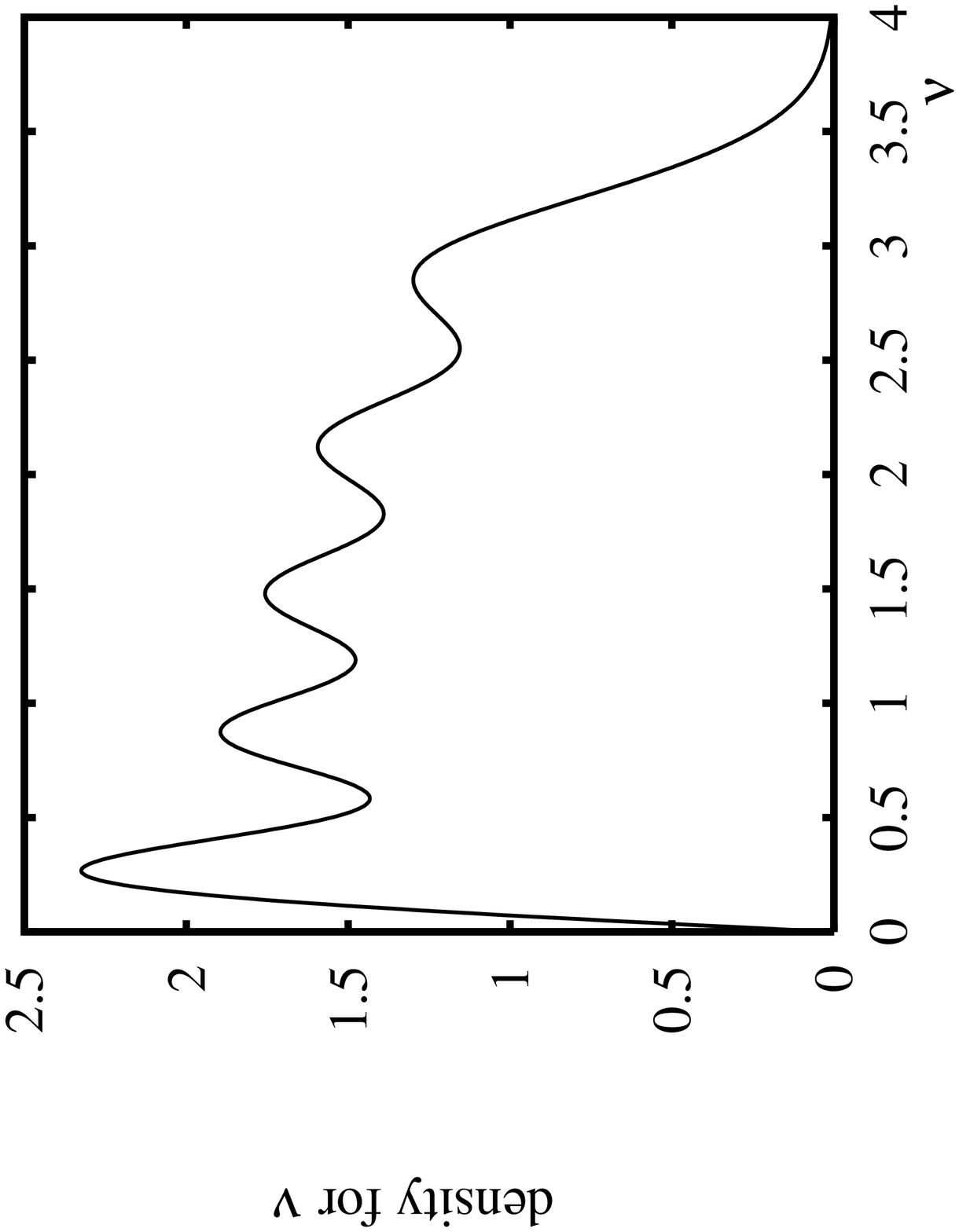}
%\phantom{aaaa}
\centerline{\LARGE Fig. 5a}
\vfill\eject
\includegraphics{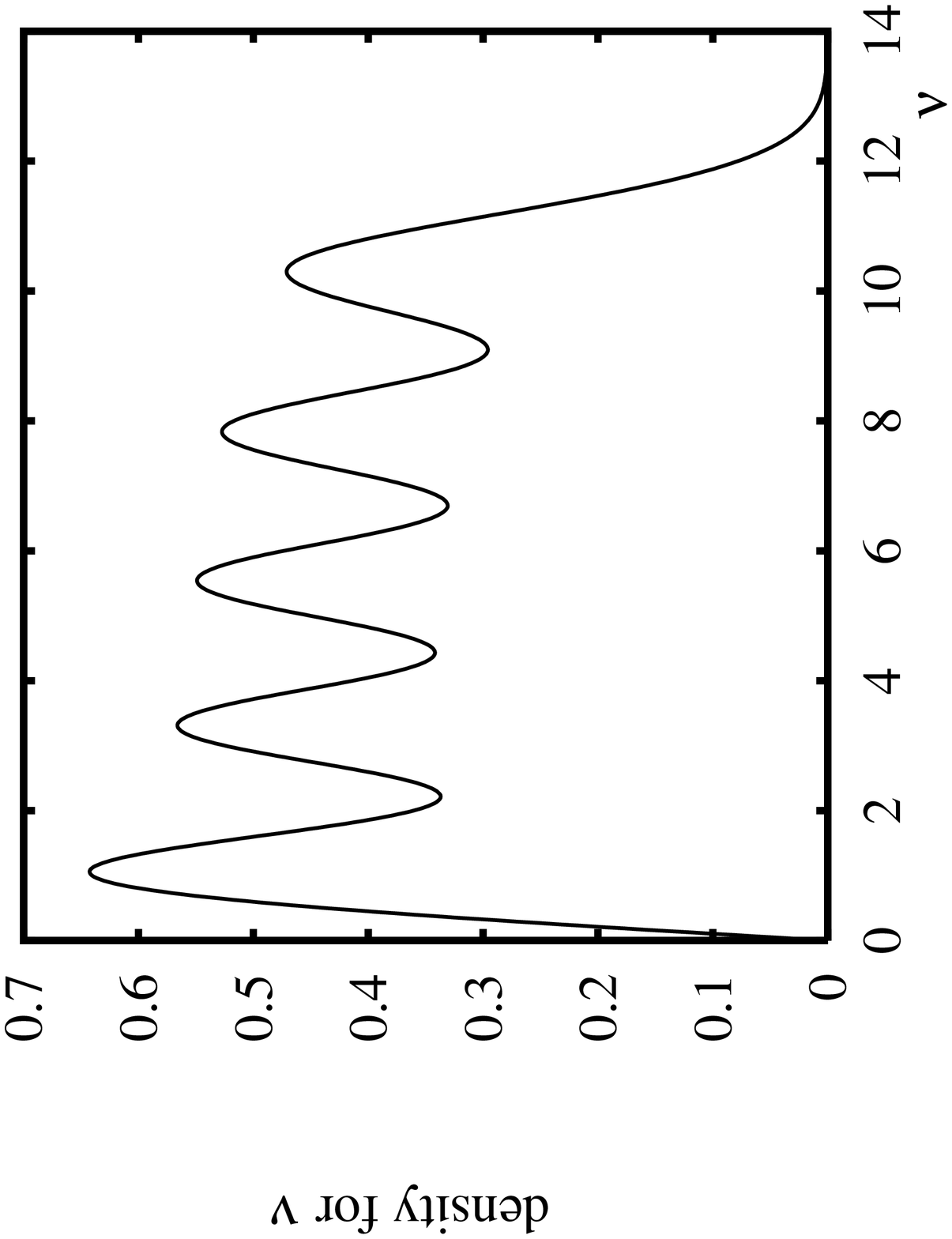}
%\phantom{aaaa}
\centerline{\LARGE Fig. 5b}
\vfill\eject
\includegraphics{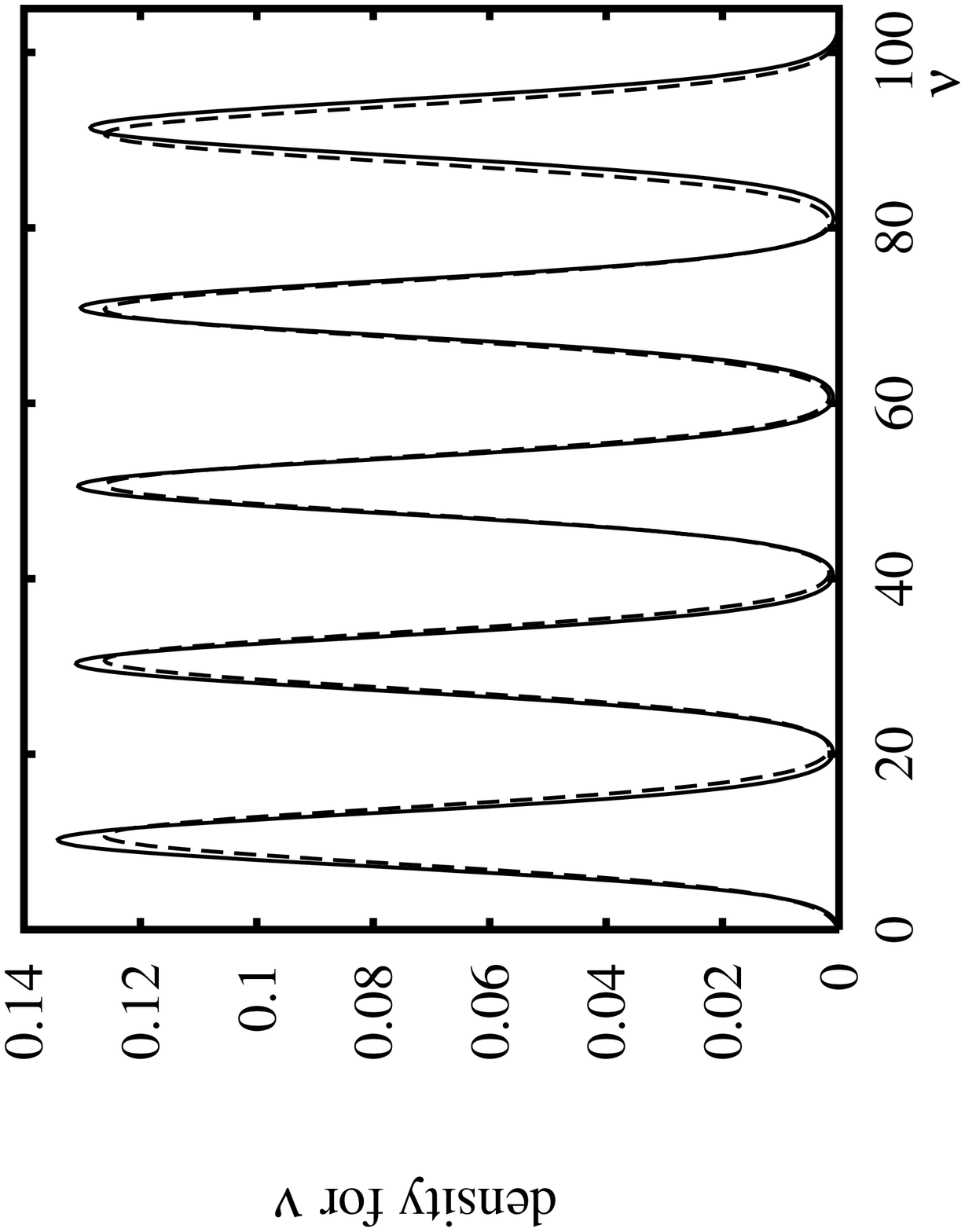}
%\phantom{aaaa}
\centerline{\LARGE Fig. 5c}
\vfill\eject

\end{document}